\documentclass[aps,pra,twocolumn,superscriptaddress]{revtex4-2}

\usepackage{dcolumn}
\usepackage{bm,bbold}
\usepackage{graphicx}
\usepackage[caption=false]{subfig}
\usepackage{float}
\usepackage{placeins}
\usepackage[squaren]{SIunits}

\usepackage{physics}
\usepackage{amsthm}
\usepackage{amsmath,bm,bbm}
\usepackage{amssymb}
\usepackage{xcolor}
\usepackage{cancel}

\usepackage[colorlinks=true, allcolors=blue]{hyperref}

\begin{document}

\title{Mechanical trisqueezing in circuit quantum acoustodynamics with SNAIL devices}

\author{Yi-Zhou Bu}
\affiliation{Institute of Physics, Chinese Academy of Science, Beijing 100190, China}
\affiliation{School of Physical Sciences, University of Chinese Academy of Sciences, Beijing 100049, China}

\author{Matteo Fadel}
\email{matteofadel@pku.edu.cn}
\affiliation{Institute of Quantum Science and Technology $\&$ Frontiers Science Center for Nano-Optoelectronics, School of Physics, Peking University, Beijing 100871, China}
\affiliation{Department of Physics, ETH Z\"{u}rich, Z\"{u}rich 8093, Switzerland}

\author{Kai Xu}
\affiliation{Institute of Physics, Chinese Academy of Science, Beijing 100190, China}
\affiliation{School of Physical Sciences, University of Chinese Academy of Sciences, Beijing 100049, China}

\author{Heng Fan}
\email{hfan@iphy.ac.cn}
\affiliation{Institute of Physics, Chinese Academy of Science, Beijing 100190, China}
\affiliation{School of Physical Sciences, University of Chinese Academy of Sciences, Beijing 100049, China}

\date{\today}

\begin{abstract}
Continuous-variable quantum information processing with mechanical oscillators requires a universal gate set, including a genuine non-Gaussian evolution. While circuit quantum acoustodynamics (cQAD) experiments have realized bilinear and quartic (four-wave mixing) operations by coupling mechanical modes to transmon qubits, the intrinsic Kerr nonlinearity of these devices precludes access to odd-order effective Hamiltonians such as trisqueezing. Here we propose a cQAD architecture in which a mechanical mode, hosted in a high-overtone bulk acoustic-wave resonator (HBAR), is coupled to a superconducting nonlinear asymmetric inductive element (SNAIL). By combining the cubic nonlinearity of the SNAIL with parametric flux driving, we engineer a three-wave mixing interaction that generates trisqueezed mechanical states, producing the characteristic triangular, star-shaped Wigner function with negative regions indicative of non-Gaussianity. We derive the effective trisqueezing Hamiltonian from the full circuit model, identify the drive frequencies and flux modulation required to isolate the desired interaction from competing terms, and quantify the achievable coupling strengths for parameters compatible with current fabrication capabilities. Our results establish a hardware-efficient route to non-Gaussian resource generation in mechanical quantum systems, extending the toolbox of cQAD toward universal continuous-variable quantum computation and enhanced quantum-limited sensing.
\end{abstract}
\maketitle

\section{Introduction}

Mechanical quantum systems have sparked great interest, thanks to their flexible design, long lifetimes, and ability to couple to many other degrees of freedom.
This also allows for the realization of hybrid quantum systems, with applications as quantum information processors~\cite{yang_mechanical_2026}, quantum memories~\cite{mechmemory}, microwave-to-optical transducers~\cite{oscarpainter}, and communication channels between remote quantum systems~\cite{dumur}.
In addition, due to their relatively large masses, mechanical quantum states are useful for sensing applications, such as the detection of forces~\cite{genuine}, gravitational waves and dark matter~\cite{omahen_ultracold_2026}, and tests of possible deviations from quantum theory~\cite{schrinski_macroscopic_2023}.

Since mechanical vibrations in the quantum regime are highly linear, they can be described as quantum harmonic oscillator modes, suitable for continuous-variable (CV) quantum information storage and processing. 
CV quantum computing is a powerful alternative to discrete-variable quantum computing based on qubits, in particular because the infinite-dimensional Hilbert space of bosonic modes allows information encoding and error correction to be performed in a hardware-efficient manner.

Crucial to CV quantum information processing and sensing applications is the availability of a universal gate set for the preparation and manipulation of arbitrary quantum states~\cite{univcv}.
Such a set includes Gaussian operations, such as displacements, phase shifts, squeezing, and beam-splitter operations, as well as at least one non-Gaussian gate. 
Previous experiments demonstrating such operations in circuit quantum acoustodynamics (cQAD) relied on coupling the mechanical modes to a superconducting transmon qubit~\cite{von_lupke_engineering_2024,fadel,kern_einstein-podolsky-rosen_2026}.
In this framework, one approach is to parametrically drive the system at carefully chosen frequencies to engineer the desired effective Hamiltonian. 
However, because the transmon nonlinearity is intrinsically dominated by its fourth-order (i.e., Kerr) term, parametrically engineered interactions have so far been limited to bilinear and quartic Hamiltonians arising from four-wave mixing processes.

\begin{figure}[t]
    \includegraphics[width=0.3\textwidth]{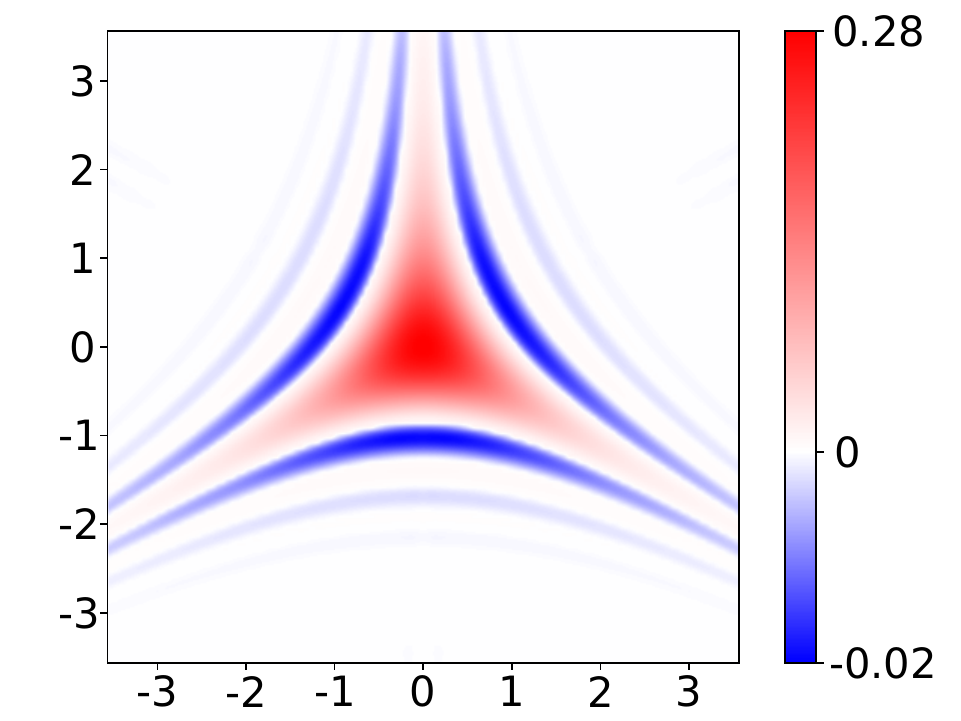}
    \caption{\textbf{Trisqueezed state.} Wigner function of an ideal trisqueezed state $|\psi\rangle=T(-0.13i)|0\rangle$, see Eq.~\eqref{trisqueeze}.}
    \label{trisqideal}
\end{figure}

A particularly interesting alternative is the engineering of three-wave mixing processes which, compared with their four-wave counterparts, can enable stronger parametric interactions, reduced calibration and control complexity, and access to odd-order effective Hamiltonians that are otherwise unavailable in Kerr-dominated architectures. One paradigmatic example is the trisqueezing (or cubic squeezing) operation
\begin{equation}
    \hat T(C)=\exp\left(C\hat c^{\dagger3}-C^*\hat c^3\right) , \label{trisqueeze}
\end{equation}
acting on a bosonic mode with annihilation operator $\hat c$, in which excitations are coherently created and annihilated in triplets. This process can thus be understood as a generalized form of squeezing~\cite{braunstein_generalized_1987}. When acting on the vacuum, it generates the state $|\psi(C)\rangle=\hat T(C)|0\rangle$, whose Wigner function has a threefold-symmetric, star-shaped distribution in phase space, see Fig.~\ref{trisqideal}. Crucially, this Wigner function takes negative values, a distinctive sign of non-classicality and non-Gaussianity.
Moreover, trisqueezed states can be converted into cubic phase states through Gaussian operations~\cite{ferraro1,ferraro2}, so that trisqueezing can serve as the non-Gaussian element of a universal gate set for CV quantum computing~\cite{univcv,hillmannprl}.
Analogously to its implementation in superconducting circuits~\cite{changprx,delsing}, realizing trisqueezing in cQAD would thus broaden the range of quantum applications accessible with mechanical systems.

In order to realize three-wave mixing in a cQAD device, one needs to couple the mechanical degree of freedom to a circuit element characterized by a cubic nonlinearity. Examples include superconducting nonlinear asymmetric inductive elements (SNAILs), asymmetrically threaded SQUIDs (ATSs), and DC-biased Josephson junctions. In this work, we consider the first option and propose a cQAD device in which a SNAIL is coupled to a high-overtone bulk acoustic-wave resonator (HBAR). Using the cubic nonlinearity of the SNAIL, operated at a Kerr-free point and activated by a parametric flux drive, we study in detail the preparation of trisqueezed states of a mechanical mode. We investigate parameter regimes that are accessible with state-of-the-art fabrication techniques, as well as the strength of the parametric processes that can be achieved.

\section{SNAIL Hamiltonian}
\label{snailsection}
Due to the small intrinsic nonlinearities, mechanical oscillators in the quantum regime do not naturally show deviations from the linear harmonic oscillator dynamics. For this reason, the engineering of multimode interactions and non-Gaussian evolutions requires the coupling to other strongly nonlinear elements. In cQAD, the latter are typically superconducting qubits, but here we consider a SNAIL~\cite{frattini}, a superconducting circuit device widely used in parametric amplifiers, frequency converters, couplers, and Kerr-cat qubits.

\begin{figure}[t]
    \includegraphics[width=0.35\textwidth]{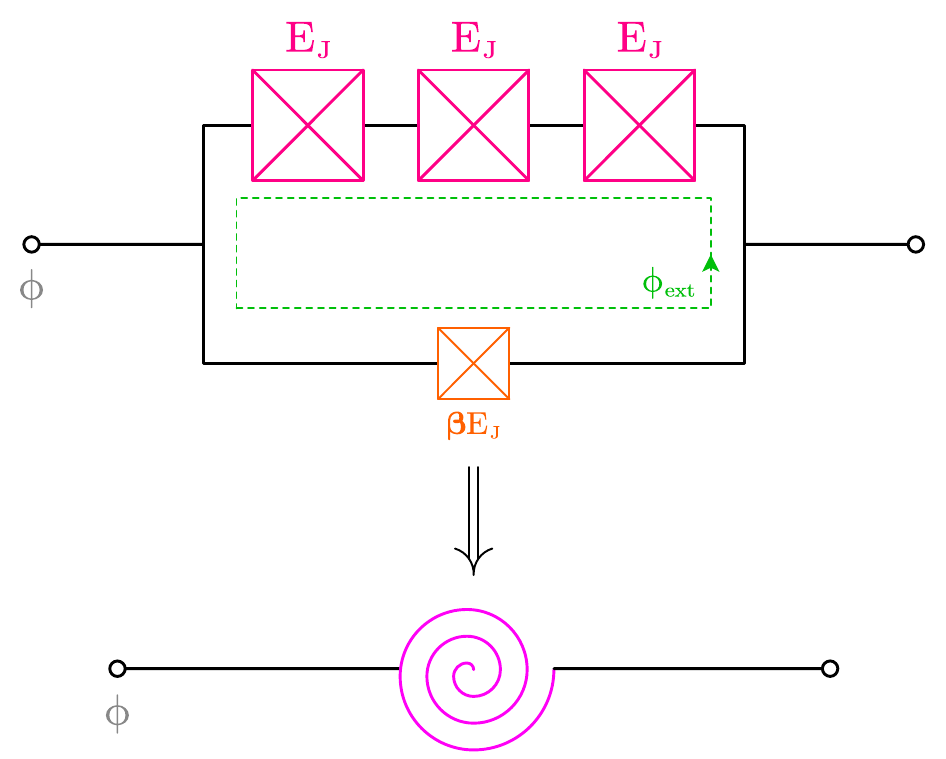}
    \caption{\textbf{Schematic of a SNAIL device and its shorthand circuit symbol.} Here, an array of $N=3$ Josephson junctions is shunted by a single junction forming a loop through which the external flux $\phi_{\text{ext}}$ is applied.}
    \label{snail}
\end{figure}

As schematically illustrated in Fig.~\ref{snail}, a SNAIL consists of an array of \(N\) Josephson junctions with Josephson energy \(E_J\), shunted by a single smaller junction of Josephson energy \(\beta E_J\), where \(\beta<1/N\) ensures a single minimum of the potential. 
These elements form a superconducting loop through which a DC magnetic flux is applied to set the operating frequency of the device, while an additional AC magnetic flux can be used for parametric modulation. 
The potential of a SNAIL can be written as:
\begin{equation}
    E(\phi,\phi_{\text{ext}}) = -\beta E_J\cos\phi-NE_J\cos\left(\frac{\phi_{\text{ext}}(t)-\phi}{N}\right),
    \label{snailinduct}
\end{equation}
where $\phi$ is the superconducting phase across the small junction, and $\phi_{\text{ext}}(t)=2\pi\Phi(t)/\Phi_0$ is the reduced external flux, with $\Phi(t)$ the applied magnetic flux and $\Phi_0=h/2e$ the flux quantum. 

The SNAIL potential has a minimum $\phi_m(\phi_{\rm ext})$ satisfying $\partial E/\partial \phi |_{\phi_m}=0$ and $\partial^2 E/\partial \phi^2 |_{\phi_m}>0$. 
Series expansion of the potential around $\phi_m$ gives:
\begin{equation}
E(\phi,\phi_{\text{ext}}) = E(\phi_m,\phi_{\text{ext}}) + \dfrac{1}{2} \left.\dfrac{\partial^2 E}{\partial\phi^2}\right\vert_{\phi_m}(\phi-\phi_m)^2 + \dots
\end{equation}
We now consider $\phi_{\text{ext}}=\phi_{\text{dc}}+\phi_{\text{ac}}(t)$, where $\phi_{\text{dc}}$ determines the bias point and $\phi_{\text{ac}}(t)$ is a weak time-dependent flux modulation. We expand the potential around the static minimum $\phi_m\equiv\phi_m(\phi_{\rm dc})$ and keep terms to first order in $\phi_{\rm ac}$, so that the modulation-induced shift of the minimum appears explicitly as a linear term in the expansion. Defining the dimensionless coefficients
\begin{equation}
c_n \equiv \frac{1}{E_J}\left.\frac{\partial^n E}{\partial\phi^n}\right|_{\phi_m,\,\phi_{\rm dc}},\qquad
c_n' \equiv \frac{1}{E_J}\left.\frac{\partial^{n+1} E}{\partial\phi^n\,\partial\phi_{\rm ext}}\right|_{\phi_m,\,\phi_{\rm dc}},
\end{equation}
where the derivative with respect to $\phi_{\rm ext}$ is taken at fixed $\phi$, the potential reads
\begin{align}
\frac{E}{E_J} &\simeq c_0 + c_0'\phi_{\rm ac}(t) + c_1'\phi_{\rm ac}(t)\,\delta\phi \nonumber\\
&\quad + \sum_{n=2}^{4}\frac{1}{n!}\left(c_n + c_n'\phi_{\rm ac}(t)\right)\delta\phi^n + \mathcal{O}(\delta\phi^5,\phi_{\rm ac}^2), \label{snailexpand}
\end{align}
with $\delta\phi\equiv\phi-\phi_m$ and $c_1 = 0$ by definition of $\phi_m$.
Here, $c_0+c_0'\phi_{\rm ac}(t)$ is a trivial energy offset that can be omitted, while the static second-order term $c_2\,\delta\phi^2/2$ determines, upon quantization, the SNAIL mode frequency. 

The linear term $c_1'\phi_{\rm ac}(t)\,\delta\phi$ describes a drive that displaces the mode: it accounts for the shift of the potential minimum induced by the flux modulation, $\delta\phi_m(t)\simeq-(c_1'/c_2)\,\phi_{\rm ac}(t)$, which is not captured by the static expansion point $\phi_m(\phi_{\rm dc})$. This term can be removed by a displacement transformation $\delta\phi\rightarrow\delta\phi+\chi\,\phi_{\rm ac}(t)$, where, for a harmonic modulation at a generic frequency $\omega_p$ ($\omega_p=\omega_d\simeq3\omega_a^0$ for the trisqueezing drive below), $\chi=-(c_1'/c_2)\,(\omega_a^0)^2/[(\omega_a^0)^2-\omega_p^2]$, with $\omega_a^0$ the SNAIL mode frequency defined below. The displacement follows the minimum adiabatically for $\omega_p\ll\omega_a^0$, is suppressed only for $\omega_p\gg\omega_a^0$, and diverges for $\omega_p\simeq\omega_a^0$, which must therefore be avoided. It renormalizes the modulated coefficients as $c_n'\rightarrow\tilde c_n'=c_n'+\chi\,c_{n+1}$, which in the adiabatic limit reduces to the total derivatives $dc_n/d\phi_{\rm dc}$. The modulated second-order term $\tilde c_2'\phi_{\rm ac}(t)\,\delta\phi^2/2$ describes a parametric modulation of the SNAIL frequency, which becomes resonant only for modulation frequencies close to twice the SNAIL frequency. The third-order term contains the desired AC nonlinearity $\tilde c_3'\phi_{\rm ac}(t)\,\delta\phi^3/6$, whose strength can be modulated through $\phi_{\rm ac}$. The third-order DC nonlinearity $c_3\,\delta\phi^3/6$, as well as higher-order terms, contribute to unwanted effects that can pose limitations to the desired dynamics. For example, Kerr-type nonlinearities can result in distortions and blockade effects limiting the maximum number of excitations in the mode. They mainly come from the native fourth-order DC term $c_4\,\delta\phi^4/24$, at first order in perturbation theory, and from the third-order DC term $c_3\,\delta\phi^3/6$, at second order through corrections to the RWA~\cite{hillmann}. The two contributions combine into an effective Kerr coefficient $K\propto c_4-5c_3^2/(3c_2)$~\cite{sivakprapplied}, while the modulated fourth-order term $c_4'\phi_{\rm ac}(t)\,\delta\phi^4/24$ gives only a small correction. Fortunately, the flexibility in the design and operation of a SNAIL device allows for minimizing these unwanted effects, for example by operating at bias fluxes where $K$ vanishes, so-called Kerr-free points (KFPs).

\begin{figure}[t]
    \includegraphics[width=0.47\textwidth]{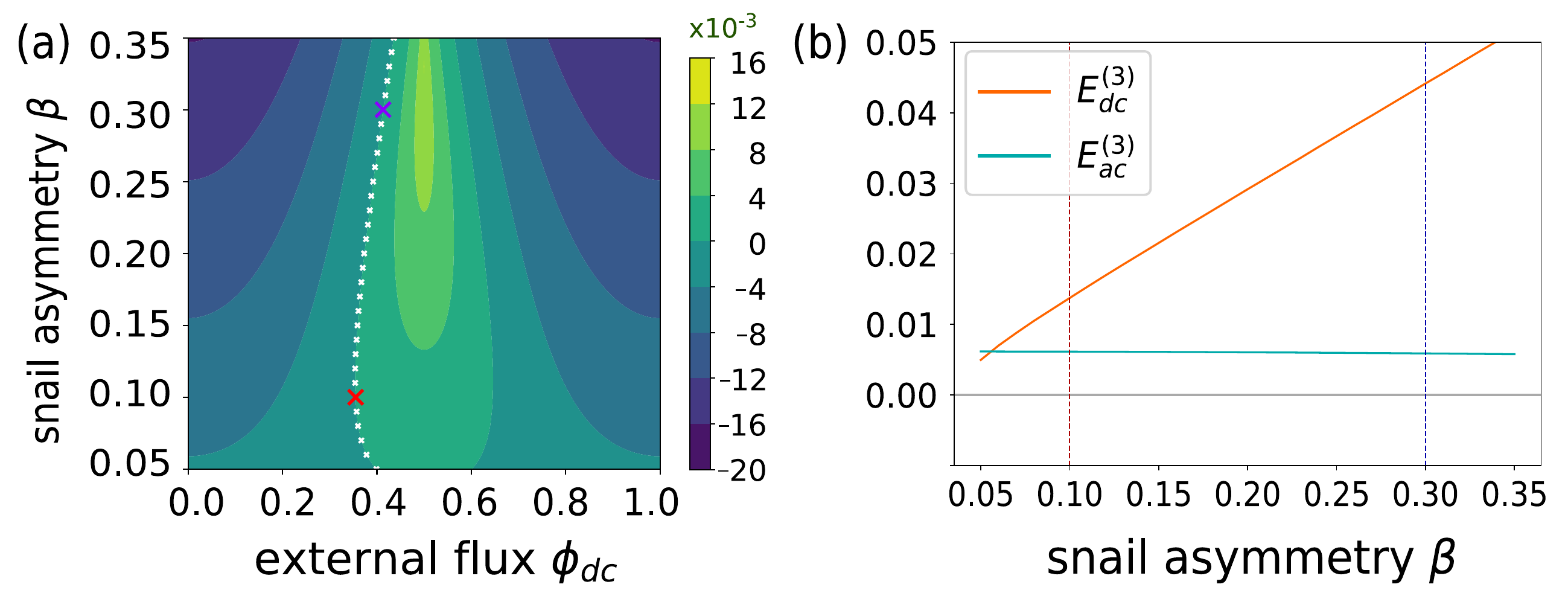}
    \caption{\textbf{Kerr-free operation of the SNAIL.} (a) Static Kerr nonlinearity $c_4-5c_3^2/(3c_2)$ of the SNAIL device considered ($N=3$), as a function of flux bias and asymmetry $\beta$. The white dotted line indicates its zero value (Kerr-free) contour. (b) Static ($c_3$) and bare modulated ($c_3'$) third-order coefficients along the Kerr-free contour, with SNAIL asymmetry $\beta$ as curve parameter. Vertical dashed lines correspond to the red ($\beta=0.1$) and blue ($\beta=0.3$) points in (a).}
    \label{snaile34}
\end{figure}

We show in Fig.~\ref{snaile34}(a) the static Kerr nonlinearity $c_4-5c_3^2/(3c_2)$ as a function of the free parameters $\Phi_{\rm dc}/\Phi_0=\phi_{\rm dc}/2\pi$ and $\beta$, for $N=3$. The white dotted line indicates where it vanishes, thus corresponding to a Kerr-free contour. Since the minimum satisfies $\beta\sin\phi_m=\sin[(\phi_{\rm dc}-\phi_m)/N]$, the static third-order coefficient can be written as $c_3=-\beta\,(1-1/N^2)\sin\phi_m$, so that its magnitude grows with the SNAIL asymmetry $\beta$ (along the Kerr-free contour, for $\beta\lesssim0.2$). In contrast, the bare modulated coefficient $c_3'=\cos[(\phi_{\rm dc}-\phi_m)/N]/N^3\simeq1/N^3$ is nearly independent of $\beta$, see Fig.~\ref{snaile34}(b). However, the renormalized coefficient $\tilde c_3'=c_3'+\chi c_4$ also depends on $\beta$ through $c_4$, which does not vanish at the KFP and grows with $\beta$. A smaller $\beta$ reduces the unwanted static $c_3$, while its effect on $\tilde c_3'$ is negligible as long as $|\chi c_4|\ll c_3'$.
In any case, $\beta$ cannot be too small, because fabricating junctions either too large or too small would be impractical. We consider in the following a value of $\beta=0.1$, compatible with existing devices~\cite{delsing,sivakprapplied}, which requires a flux bias $\Phi_{\rm dc}\simeq0.399\,\Phi_0$ ($\phi_{\rm dc}\simeq2.51$) to operate at the KFP, where $c_2\simeq0.267$, $c_3\simeq-0.067$, $c_4\simeq0.028$ and $c_3'\simeq0.037$, while in the adiabatic limit $\chi\simeq1.24$ and $\tilde c_3'\simeq0.072$.

Taking into consideration the charging energy $E_C=e^2/2C$, with $C$ the total capacitance shunting the SNAIL, the SNAIL Hamiltonian reads $H=4E_C\hat n^2+E(\hat\phi,\phi_{\rm ext})$, with $[\hat\phi,\hat n]=i$. It is possible to quantize it around the minimum as $\delta\phi\rightarrow\varphi_{\mathrm{zpf}}(\hat a_0^\dagger+\hat a_0)$ and $\hat n\rightarrow i(\hat a_0^\dagger-\hat a_0)/(2\varphi_{\mathrm{zpf}})$, with $\varphi_{\mathrm{zpf}}=\left[2E_C/(E_Jc_2)\right]^{1/4}$ chosen to diagonalize the static quadratic part. Omitting constant offsets and setting $\hbar=1$, we obtain the Hamiltonian:
\begin{align}
H_{\text{SNAIL}} &= \omega_a^0 \hat{a}_0^\dagger \hat{a}_0
+ \sum_{n=1}^{4} g^{(n)} \left(\hat{a}_0^\dagger + \hat{a}_0\right)^n,\nonumber\\
g^{(n)}&\equiv g^{(n)}_{\text{dc}}+g^{(n)}_{\text{ac}}\phi_{\text{ac}}(t),
\label{snailnormalhamiltonian}
\end{align}
with bare-mode annihilation operator $\hat{a}_0$, bare-mode frequency $\omega_a^0=\sqrt{8E_C E_J c_2}$, and coefficients
\begin{equation}
g^{(n)}_{\text{dc}}=\frac{E_J c_n\,\varphi_{\mathrm{zpf}}^n}{n!},\qquad
g^{(n)}_{\text{ac}}=\frac{E_J c_n'\,\varphi_{\mathrm{zpf}}^n}{n!},
\label{snailcouplings}
\end{equation}
where $g^{(1)}_{\text{dc}}=0$ since $c_1=0$, and $g^{(2)}_{\text{dc}}\equiv0$ since the static quadratic term is already included in $\omega_a^0$. In terms of these parameters, the Kerr coefficient introduced above, defined through $H_{\rm Kerr}=(K/2)\,\hat a_0^{\dagger2}\hat a_0^2$, reads $K=12g^{(4)}_{\text{dc}}-60\,(g^{(3)}_{\text{dc}})^2/\omega_a^0=(E_C/c_2)\left[c_4-5c_3^2/(3c_2)\right]$, which vanishes at the KFP. If the drive term $g^{(1)}$ is eliminated by the displacement discussed above, the modulated coefficients are replaced by $\tilde g^{(n)}_{\text{ac}}=E_J\tilde c_n'\varphi_{\mathrm{zpf}}^n/n!$. Details of this derivation are in Appendix~\ref{appendix:quant}.

\section{HBAR and piezoelectric coupling}
The mechanical degree of freedom considered in this work is an acoustic vibration mode hosted inside an HBAR, which can reach energy relaxation times exceeding $T_1=400\,\mu\mathrm{s}$~\cite{belles_loss_2026}.
These long lifetimes result from the fact that HBAR modes live inside high-quality crystalline substrates, and make them especially suited for quantum information storage, processing, and sensing applications.

A typical HBAR consists of a bulk substrate (e.g., sapphire) acting as a Fabry--P\'erot resonator for acoustic vibrations, which are excited through a disk of piezoelectric material (e.g., aluminum nitride).
The latter allows for a bidirectional coherent interaction between the vibration mode and the electric field generated by, e.g., a superconducting circuit device.
Because of the native piezoelectric coupling, the interaction Hamiltonian takes, within the rotating-wave approximation, the beam-splitter form
\begin{equation}
    \hat H_{\rm int} = g\left(\hat a_0^\dagger \hat c_0+\hat a_0\hat c_0^\dagger\right) ,
\end{equation}
where $g$ is the coupling rate, and $\hat a_0$ and $\hat c_0$ are the annihilation operators of the SNAIL mode and of the mechanical mode, associated with the electric and strain field, respectively. The rotating-wave approximation is valid for $g\ll\omega_a^0,\omega_c$, with $\omega_c$ the mechanical mode frequency.

\section{Parametric Pump Scheme}
In this section we begin from the full HBAR-SNAIL system Hamiltonian and derive an effective three-phonon squeezing interaction from an appropriate parametric modulation of the SNAIL flux.
A detailed version of this derivation can be found in Appendix~\ref{appendix:trisq}.

We start from the Hamiltonian in the frame rotating at the phonon frequency $\omega_c$ for both modes, which reads
\begin{align}
    \hat H &= \Delta_a\hat a_0^\dagger \hat a_0+g\left(\hat a_0^\dagger \hat c_0+\hat a_0\hat c_0^\dagger\right)\nonumber\\
    &\quad+\Omega_3\cos(\omega_d t)\left(\hat a_0^\dagger e^{i\omega_c t}+\hat a_0 e^{-i\omega_c t}\right)^3 , \label{hsnail}
\end{align}
where $\Delta_a=\omega_a^0-\omega_c$, $\omega_d$ is the modulation frequency of the SNAIL flux, $\phi_{\text{ac}}(t)=\phi_{\text{ac}}\cos(\omega_d t)$, and $\Omega_3\equiv\tilde g^{(3)}_{\text{ac}}\phi_{\text{ac}}$.
Here, we assume operation at the KFP $\beta=0.1$, $\Phi_{\rm dc}\simeq0.399\,\Phi_0$, where the Kerr coefficient $K$ vanishes. At leading order, the static shift of the SNAIL frequency, $12g^{(4)}_{\text{dc}}-60\,(g^{(3)}_{\text{dc}})^2/\omega_a^0$, coincides with $K$ and thus vanishes as well, so that $\omega_a^0$ is the SNAIL frequency. The remaining static terms $g^{(3)}_{\text{dc}}$ and $g^{(4)}_{\text{dc}}$ are non-resonant and are dropped within the RWA. Among the modulated terms, those with an even number of operators, $g^{(2)}_{\text{ac}}$ and $g^{(4)}_{\text{ac}}$, cannot be resonant with $\omega_d\simeq3\omega_c$ and are dropped as well. The non-resonant linear drive $g^{(1)}_{\text{ac}}$ is removed by the displacement transformation discussed in Sec.~\ref{snailsection}, which renormalizes $g^{(3)}_{\text{ac}}\rightarrow\tilde g^{(3)}_{\text{ac}}$. For $\omega_d\simeq3\omega_a^0$, one finds $\chi\simeq-0.16$ and $\tilde c_3'\simeq0.033$. Equation~\eqref{hsnail} thus keeps only the modulated third-order term, which is activated by the flux modulation.

We now assume the dispersive regime $|g/\Delta_a| \ll 1$, which allows us to remove the beam-splitter coupling through a Schrieffer--Wolff transformation. To leading order in $g/\Delta_a$, this hybridizes the bare modes as $\hat a_0\simeq\hat a-(g/\Delta_a)\,\hat c$ and $\hat c_0\simeq\hat c+(g/\Delta_a)\,\hat a$, where $\hat a$ ($\hat c$) annihilates the SNAIL-like (phonon-like) dressed mode, resulting in
\begin{align}
    \hat H&=\Delta_a\hat a^\dagger \hat a+\chi_{\rm L}\left(\hat a^\dagger \hat a-\hat c^\dagger \hat c\right)+\Omega_3\cos(\omega_d t)\nonumber\\
    &\quad\times\left(\hat a^\dagger e^{i\omega_c t}-\frac{g}{\Delta_a}\hat c^\dagger e^{i\omega_c t}+\hat a e^{-i\omega_c t}-\frac{g}{\Delta_a}\hat c e^{-i\omega_c t}\right)^3 . \label{hsnailswt}
\end{align}
Another rotating frame absorbs the Lamb shift $\chi_{\rm L}=g^2/\Delta_a$ of the phonon-like mode as
\begin{align}
   \hat H&=(\Delta_a+\chi_{\rm L})\hat a^\dagger \hat a+\frac{\Omega_3}{2}\left(e^{i\omega_d t}+e^{-i\omega_d t}\right)\nonumber\\
   &\quad\times\left(\hat a^\dagger e^{i\omega_c t}-\frac{g}{\Delta_a}\hat c^\dagger e^{i(\omega_c-\chi_{\rm L})t}+\text{h.c.}\right)^3 . \label{hsnaillamb}
\end{align}
If one now modulates the SNAIL flux at three times the shifted phonon frequency, $\omega_d=3(\omega_c-\chi_{\rm L})$, the down-conversion of one modulation quantum into three phonons becomes resonant. Provided that $\Omega_3\ll|\Delta_a|$, all terms involving the SNAIL-like mode remain detuned by multiples of $\Delta_a$. The leading neglected ones are $\hat a^\dagger\hat c^{\dagger2}$ and $\hat a^{\dagger2}\hat c^\dagger$, detuned by $\simeq\Delta_a$ and $2\Delta_a$, together with their Hermitian conjugates. Keeping only the resonant term, the cubic interaction results in the effective trisqueezing Hamiltonian
\begin{align}
    \hat H_{\rm tsq}&=-\frac{\Omega_3}{2}\left(\frac{g}{\Delta_a}\right)^3\left(\hat c^{\dagger3}+\hat c^3\right) , \label{trisqsnailhbarham}\\
    \Omega_3&=\frac{1}{6}\,E_J\,\tilde c_3'\,\varphi_{\mathrm{zpf}}^3\,\phi_{\text{ac}} . \label{trisqsnailhbarg3}
\end{align}
Equation~\eqref{trisqsnailhbarham} shows how SNAIL-HBAR coupled systems can, in principle, show the effect of three-phonon squeezing. 
It is important to remark that our derivation relies on the RWA and on the dispersive approximation, which allowed us to drop non-resonant terms and higher-order corrections in $g/\Delta_a$. 
For the parameters used below, $|g/\Delta_a|\simeq0.25$ and $\Omega_3$ is not much smaller than $|\Delta_a|$, so these approximations are only marginally satisfied; the numerical simulations therefore retain all neglected terms.
In the following sections we demonstrate this effective dynamics through numerical simulations, and characterize the final states obtained. These simulations are based on the full system Hamiltonian, such that the importance of the terms dropped throughout our derivation can be estimated. 
As a side remark, we note that alternative derivations based on Floquet theory can also be applied.

\section{Trisqueezing Metrics}
Unlike normal squeezing, where a decrease in uncertainty only happens along one phase-space direction, trisqueezing results in non-Gaussian states like the one shown in Fig.~\ref{trisqideal}, for which the variance alone provides an insufficient metric.
In fact, since an ideal trisqueezed state only populates Fock states whose number is a multiple of three, $\langle\hat c\rangle=\langle\hat c^2\rangle=0$, and its variance is the same along all phase-space directions. For this reason, dedicated nonlinear squeezing measures have been developed in the literature.

First, since Eq.~\eqref{trisqsnailhbarham} is only an approximation of the actual dynamics, the fidelity with respect to an ideal trisqueezed state is not an appropriate figure of merit here. 
One alternative quantity that is also of immediate practical interest for sensing applications is the quantum Fisher information (QFI)
\begin{equation}
    F[\rho,\hat O]=2\sum_{\lambda_k+\lambda_l>0}\frac{(\lambda_k-\lambda_l)^2}{\lambda_k+\lambda_l}|\langle k|\hat O|l\rangle|^2 , \label{fisher}
\end{equation}
where $\lambda_k$ are the eigenvalues of $\rho$ associated with the eigenvectors $|k\rangle$.
Eq.~\eqref{fisher} characterizes the sensitivity of a generic quantum state $\rho$ to a perturbation generated unitarily by the operator $\hat O$.
Here, a natural choice is to consider the sensing of rotations in phase space, such that $\hat O = \hat c^\dagger \hat c$~\cite{fadel_quantum_2025}.

Another choice of metric for trisqueezing is a classical quantity that generalizes the variance to reflect the asymmetry of a distribution, and for this reason is usually called ``skewness''. 
Concretely, it measures the third central moment of the distribution of a quadrature $\hat y_\theta=\hat x\cos\theta+\hat p\sin\theta$~\cite{changprx}, with $\hat x=(\hat c+\hat c^\dagger)/\sqrt{2}$ and $\hat p=-i(\hat c-\hat c^\dagger)/\sqrt{2}$, as
\begin{equation}
    S(\rho,\theta)=\left\langle\left(\hat y_\theta-\langle \hat y_\theta\rangle\right)^3\right\rangle \;.\label{thirdmoment}
\end{equation}
Note that in Eq.~\eqref{thirdmoment} we left out the normalization by $\sigma^3$, with $\sigma$ the standard deviation, which is present in most definitions. This is because we want $S(\rho,\theta)$ to increase monotonically with the trisqueezing parameter $C$, at least for an ideal state. 
We represent the skewness $S(\rho,\theta)$ in polar coordinates as a function of $\theta$.

\begin{figure}[t]
    \includegraphics[width=0.45\textwidth]{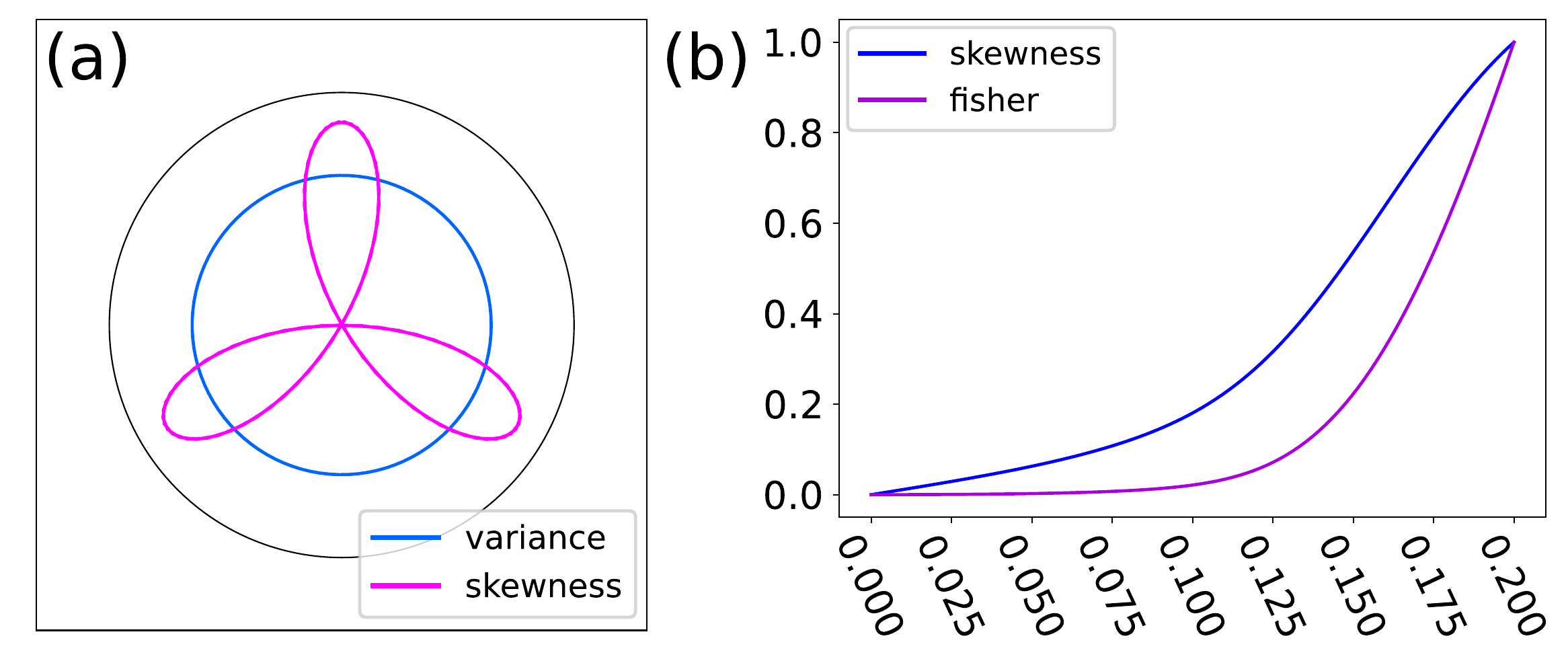}
    \caption{\textbf{Trisqueezing metrics.} (a) Standard deviation $\sigma(\rho,\theta)$ (blue) and skewness (pink) of an ideally trisqueezed state $|\psi\rangle=\hat T(-0.13i)|0\rangle$. (b) QFI (purple) and skewness (blue), as a function of the trisqueeze parameter C in Eq.~\eqref{trisqueeze}. The curves have been normalized to their maximum value.}
    \label{skewqficompare}
\end{figure}

As shown in Fig.~\ref{skewqficompare}(a), the skewness $S(\rho,\theta)$ of an ideal trisqueezed state has the same threefold rotational symmetry as its Wigner function. We therefore define the skewness of a state as its largest value along all directions,
\begin{equation}
    S(\rho)\equiv\max_\theta S(\rho,\theta) \;. \label{skewdef}
\end{equation}
Fig.~\ref{skewqficompare} shows both the skewness $S(\rho)$ and the QFI for an ideal trisqueezed state $|\psi\rangle=\hat T(C)|0\rangle$, as a function of $C$. Both metrics grow monotonically with $C$. Note, however, that the ideal trisqueezing operator is not well defined in the infinite-dimensional Fock space~\cite{fisher_impossibility_1984,braunstein_generalized_1987}, so that $\hat T(C)|0\rangle$ is computed in a Fock space truncated at $N_{\max}$ phonons. The results are independent of $N_{\max}$ only for $C\lesssim0.1$, while the steep increase of both metrics at larger $C$ depends on the truncation.
As the qualitative behavior of these two metrics is similar, in the following we will use as a metric the skewness only.

\section{Numerical Simulations}
In this section we present the results of our numerical simulations of the trisqueezing dynamics, the simulations are performed using QuTiP~\cite{github}.
Although the derivation of Eq.~\eqref{trisqsnailhbarham} starts from the Hamiltonian Eq.~\eqref{hsnail}, which only keeps third order AC term,  for our simulations we consider every term up to 4th order, as in Eq.~\eqref{snailexpand}. 
Table~\ref{snailtable} summarizes all nonlinear terms included and their values. 
Note that at the KFP $g^{(4)}_{\rm dc}$ does not vanish, but compensates the Kerr nonlinearity induced at second order by $g^{(3)}_{\rm dc}$, i.e., $12g^{(4)}_{\rm dc}\simeq60(g^{(3)}_{\rm dc})^2/\omega_a^0$.
In addition, we fix the coupling $g=2\pi\cdot\unit{350}{kHz}$, the SNAIL center frequency to $\sim\unit{6.5}{GHz}$, and the HBAR's mode detuning $\Delta=-\Delta_a$ to be around \unit{1-2}{MHz}.
The amplitude of the flux modulation $\phi_{\text{ac}}$ is slowly (adiabatically) ramped up to its maximum value in \unit{0.2}{\mu s}. 

\begin{table}[t]
\caption{\textbf{Simulation parameters.} SNAIL nonlinearity values for numerical simulations. AC entries include the flux amplitude, evaluated at $\phi_{\rm ac}$.}
\label{snailtable}
\begin{ruledtabular}
\begin{tabular}{lcr}%\unit{}{}
2nd order DC&$g^{(2)}_{\rm dc}/2\pi$& \unit{1625}{MHz}\\
2nd order AC&$g^{(2)}_{\rm ac}/2\pi$& \unit{-35.6}{MHz}\\
3rd order DC&$g^{(3)}_{\rm dc}/2\pi$& \unit{-17.8}{MHz}\\
3rd order AC&$g^{(3)}_{\rm ac}/2\pi$& \unit{4.99}{MHz}\\
4th order DC&$g^{(4)}_{\rm dc}/2\pi$& \unit{-1.52}{kHz}\\
4th order AC&$g^{(4)}_{\rm ac}/2\pi$& \unit{4.54}{kHz}\\
\end{tabular}
\end{ruledtabular}
\end{table}

\begin{figure}[t]
    \centering
    \includegraphics[width=0.47\textwidth]{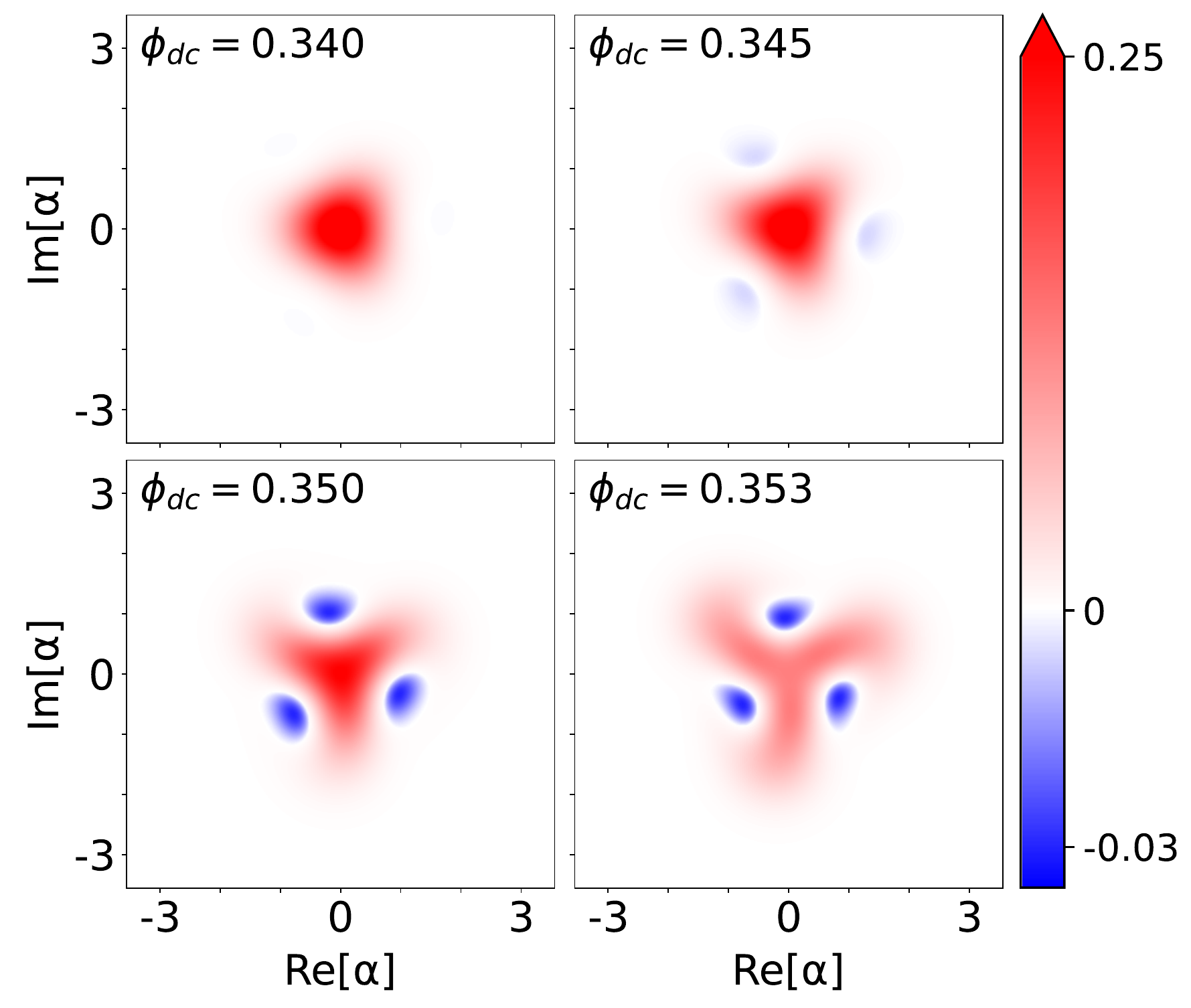}
    \caption{\textbf{Trisqueezing simulations.} Wigner function of the HBAR state for four flux biases $\Phi_{\rm dc}/\Phi_0$. Values are clipped to the range $[-0.03,0.25]$.}\label{wignercomp}
\end{figure}

In our numerical simulations we have investigated whether changing certain parameters will affect the amount of trisqueezing obtained, using skewness as the metric.
The first parameter we have considered is the SNAIL flux bias $\Phi_{\rm dc}$. Figure~\ref{wignercomp} shows the final HBAR states after \unit{0.5}{\mu s} of parametric driving, for different values of $\Phi_{\rm dc}$.
We can observe three-fold symmetric features in all four subplots, supporting the validity of our scheme, although with artifacts compared to ideal trisqueezed states likely due to the residual Kerr effect.

\begin{figure}[b]
    \centering
    \includegraphics[width=0.45\textwidth]{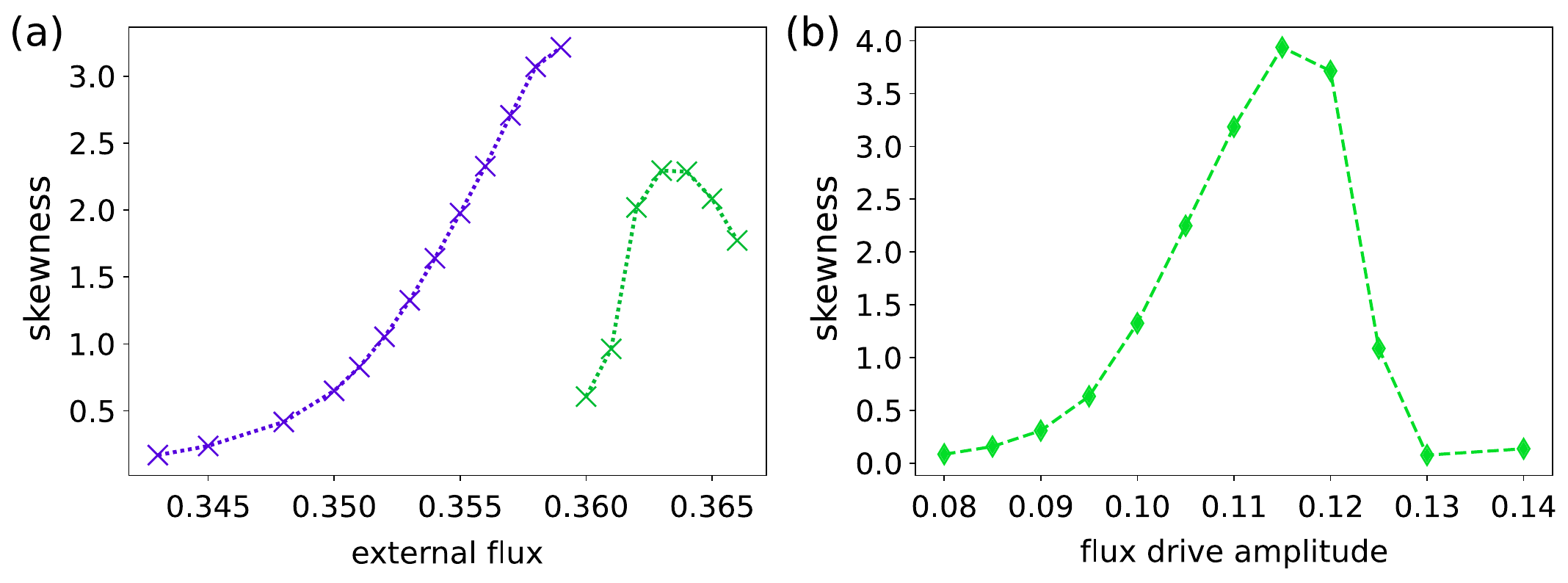}
    \caption{\textbf{Skewness vs. flux bias and drive amplitude.} Evolution time set to \unit{0.5}{\mu s} and SNAIL's red detuning \unit{1.4}{MHz}. (a) Drive amplitude fixed at $\phi_{\rm ac}=0.1$: the skewness jumps around $\Phi_{\rm dc}/\Phi_0\simeq0.36$. (b) Flux bias fixed at $\Phi_{\rm dc}/\Phi_0=0.353$: the skewness drops as $\phi_{\rm ac}$ approaches 0.12.}
    \label{skewcomp}
\end{figure}
{%\color{cyan}
In Fig.~\ref{skewcomp}(a) we show the skewness of the resulting state for $\Phi_{\rm dc}/\Phi_0\in[0.343,0.366]$, where a discontinuity around 0.36 can be observed.
%The discontinuity around 0.36 appears due to the Hilbert space truncations in the simulation, which results in an upper limit for the skewness.
In addition, we plot in Fig.~\ref{skewcomp}(b) the effect of changing the flux drive amplitude $\phi_{\text{ac}}$,
note that above $\phi_{\text{ac}}\simeq0.12$ the skewness decreases significantly, close to the perturbative estimate of Appendix~\ref{appendix:trans}.
This shows that, even if we assure adiabatic pumping, $\phi_{\text{ac}}$ still cannot exceed a certain threshold.
Both these phenomena are related to a change in Floquet states when passing through anti-crossings. Discussions on this topic can be found in Appendix~\ref{appendix:trans}.
}

So far, losses have not been considered in our simulations, due to memory limitations. On the other hand, however, we have observed that the relevant trisqueezing dynamics should last on the order of a few microseconds, which is considerably shorter than HBAR's mode lifetime. Thus, we do not expect decoherence mechanisms to contribute significantly to the final result. This is also supported by prior numerical works on similar topics, where introducing losses only causes a 0.2\% drop to a 97.4\% fidelity when generating cubic phase states~\cite{hillmannprl}. This implies that unwanted nonlinearities make up for the majority of the infidelity.

\section{Circuit Configuration}
\begin{figure}[t]
    \includegraphics[width=0.45\textwidth]{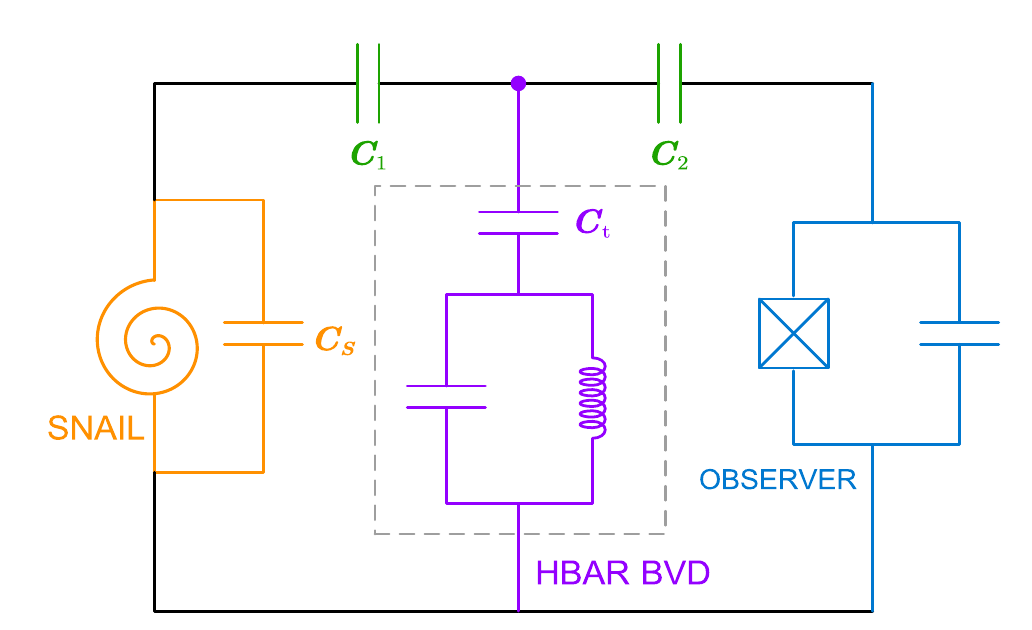}
    \caption{\textbf{Circuit implementation of the SNAIL-HBAR system.} The HBAR (purple) is represented by a parallel BVD equivalent circuit, and the capacitors $C_1$ and $C_2$ (green) couple the SNAIL and the transmon to the HBAR, respectively.}
    \label{cqedthing}
\end{figure}

We are now going to relate our ideal model to a concrete circuit configuration and verify the performance of our scheme for realistic parameters. The circuit model we consider is shown in Fig.~\ref{cqedthing}. There, besides the HBAR and SNAIL devices, a spectator transmon qubit (blue) is included to read out the mechanical state, e.g., through direct Wigner function measurement~\cite{parity}.
\begin{figure}[t]
    \includegraphics[width=0.35\textwidth]{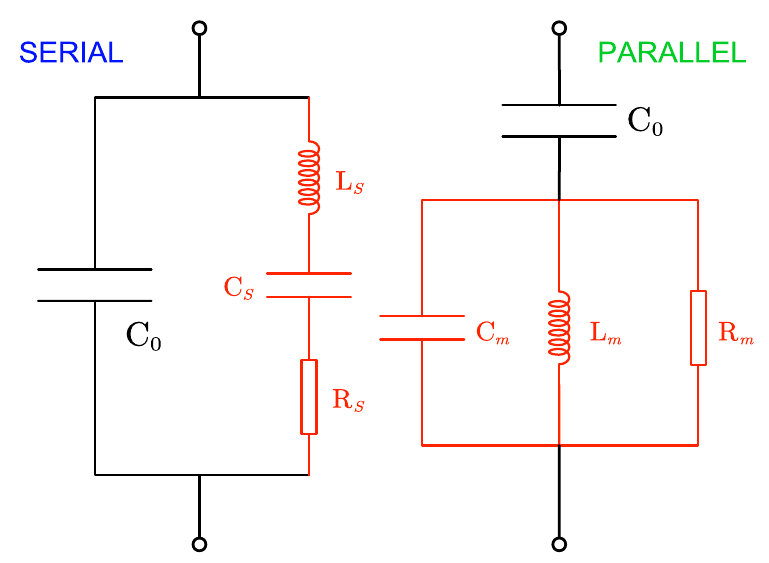}
    \caption{\textbf{Comparison between series and parallel BVD HBAR-equivalent circuits.} Resistors, describing losses, are omitted in the circuit QED model.}
    \label{seriepara}
\end{figure}
\begin{table}[b]
\caption{\textbf{Circuit parameters.}
Parameters for the circuit diagram in Fig.~\ref{cqedthing}, expressed in frequency units as $E(C)=e^2/2hC$ for capacitances and $E(L)=\hbar/8\pi e^2L$ for inductances, together with the resulting coupling strengths. All values are obtained by dividing frequencies and couplings by $2\pi$.
}\label{cqedtable}
\begin{ruledtabular}
\begin{tabular}{lcr}
SNAIL capacitance&$E(C_s)$&\unit{24.2}{MHz}\\%800 fF
Qubit capacitance&$E(C_q)$&\unit{161.4}{MHz}\\%120 fF
SNAIL-HBAR coupling capacitance&$E(C_1)$&\unit{1.3}{GHz}\\%15 fF
Qubit-HBAR coupling capacitance&$E(C_2)$&\unit{3.9}{GHz}\\%5 fF
HBAR series capacitance&$E(C_t)$&\unit{9.7}{GHz}\\%2 fF
Parallel mechanical capacitance&$E(C_m)$&\unit{48.4}{kHz}\\%400 pF
Parallel mechanical inductance&$E(L_m)$&\unit{1.09\times10^5}{GHz}\\%1.5 pH
SNAIL-qubit coupling&$g_{\rm sq}$&\unit{49.7}{MHz}\\
SNAIL-HBAR coupling&$g_{\rm sm}$&\unit{0.31}{MHz}\\
Qubit-HBAR coupling&$g_{\rm qm}$&\unit{0.24}{MHz}\\
\end{tabular}
\end{ruledtabular}
\end{table}

The mechanical branch (purple) is represented by a parallel Butterworth-Van Dyke (BVD) circuit~\cite{morgan,arnau}. Its components are related to those of the usual series BVD circuit, consisting of the static capacitance $C_0$ in parallel with the motional elements $L_{\rm ser}$ and $C_{\rm ser}$ (see Fig.~\ref{seriepara}), as
\begin{equation}
    C_m=\frac{1+\gamma}{\gamma}\,C_0,\qquad L_m=\frac{\gamma}{(1+\gamma)^2\,\omega_{\rm ser}^2C_0},\label{satzingertransform}
\end{equation}
where $\omega_{\rm ser}=1/\sqrt{L_{\rm ser}C_{\rm ser}}$ is the series resonance frequency, and the parallel $L_mC_m$ circuit is connected in series with a capacitance $(1+\gamma)C_0\simeq C_0$. In the above equation, $\gamma=C_{\rm ser}/C_0$ is proportional to the piezoelectric coupling coefficient $k^2_{\rm eff}$ as~\cite{oconnell}
\begin{equation}
    \gamma=\frac{C_{\rm ser}}{C_0}=\frac{8}{\pi^2}k^2_{\rm eff}.\label{keff}
\end{equation}
Here, $C_0$ stands for the electrical capacitance of the HBAR resonator. For HBARs, $k^2_{\rm eff}$ usually takes values around $2\times10^{-4}$, which for the parameters of Table~\ref{cqedtable} gives $C_0\approx\unit{65}{fF}$ and a series mechanical capacitance $C_{\rm ser}\approx\unit{0.01}{fF}$. The capacitor $C_t$ in Fig.~\ref{cqedthing} comprises two capacitors in series: one being $C_0$, the other being a much smaller $C_d$ representing the HBAR's coupling to the outside, potentially due to its antenna structure. As these two are in series, $C_d$ dominates and $C_t\approx C_d$. To reduce the qubit-HBAR coupling below \unit{1}{MHz}, $C_d$ should be as small as 1-2\,fF.

Quantizing the circuit using the standard circuit QED approach~\cite{blackbox} and applying the RWA gives the Hamiltonian 
\begin{align}
    \hat H&=\hat H_0+g_{\rm sq}\left(\hat a_0^\dagger \hat q+\hat a_0\hat q^\dagger\right)\nonumber\\
    &\quad+g_{\rm qm}\left(\hat q^\dagger \hat c_0+\hat q\hat c_0^\dagger\right)+g_{\rm sm}\left(\hat a_0^\dagger \hat c_0+\hat a_0\hat c_0^\dagger\right) ,\label{cqed1}
\end{align}
which describes three coupled modes, with annihilation operators $\hat a_0$, $\hat q$, and $\hat c_0$ for the SNAIL, the transmon qubit, and the HBAR, respectively. Here, $\hat H_0=\omega_a\hat a_0^\dagger\hat a_0+\omega_q\hat q^\dagger\hat q+\omega_c\hat c_0^\dagger\hat c_0$, with $\omega_a$ the flux-dependent SNAIL frequency ($\omega_a^0$ above), plus the SNAIL nonlinear terms of Eq.~\eqref{snailnormalhamiltonian} and the transmon anharmonicity, and $g_{\rm sm}$ corresponds to the coupling $g$ of the previous sections. Derivation details can be found in Appendix~\ref{appendix:cqed}.
Table~\ref{cqedtable} shows a set of realistic experimental parameters, as well as the three resulting coupling strengths. In particular, $g_{\rm sm}=\unit{0.31}{MHz}$ is close to the value used in our simulations (\unit{0.35}{MHz}), and well within the experimentally accessible range.

From Table~\ref{cqedtable} we note that, although two capacitors separate the SNAIL and the spectator qubit, their electrical coupling $g_{\rm sq}$ is still two orders of magnitude larger than the electromechanical couplings $g_{\rm sm}$ and $g_{\rm qm}$. 
As a result, both the SNAIL and the spectator qubit acquire significant frequency shifts, comparable to the SNAIL-HBAR detuning. Assuming the spectator qubit to be tuned to \unit{10}{GHz} while trisqueezing takes place near \unit{6.5}{GHz} gives a SNAIL frequency shift $\chi_{\rm sq}=g_{\rm sq}^2/\Delta_{\rm sq}\approx2\pi\cdot\unit{0.7}{MHz}$, with $\Delta_{\rm sq}=\omega_q-\omega_a$. 
This shift should be taken into account from the start when parametrically engineering the trisqueezing dynamics: in Eq.~\eqref{hsnail}, $\Delta_a$ should already include it through $\omega_a^0\rightarrow\omega_a^0-\chi_{\rm sq}$.

\section{SNAIL frequency tunability}
In our proposed device, tuning the interactions between circuit elements relies on the frequency tunability of the SNAIL and of the spectator qubit, e.g., through an external flux bias or the ac Stark shift. Alternatively, one could consider tunable couplers.
To perform our measurement scheme, two main aspects need to be considered.

\begin{figure}[b]
    \centering
    \includegraphics[width=0.46\textwidth]{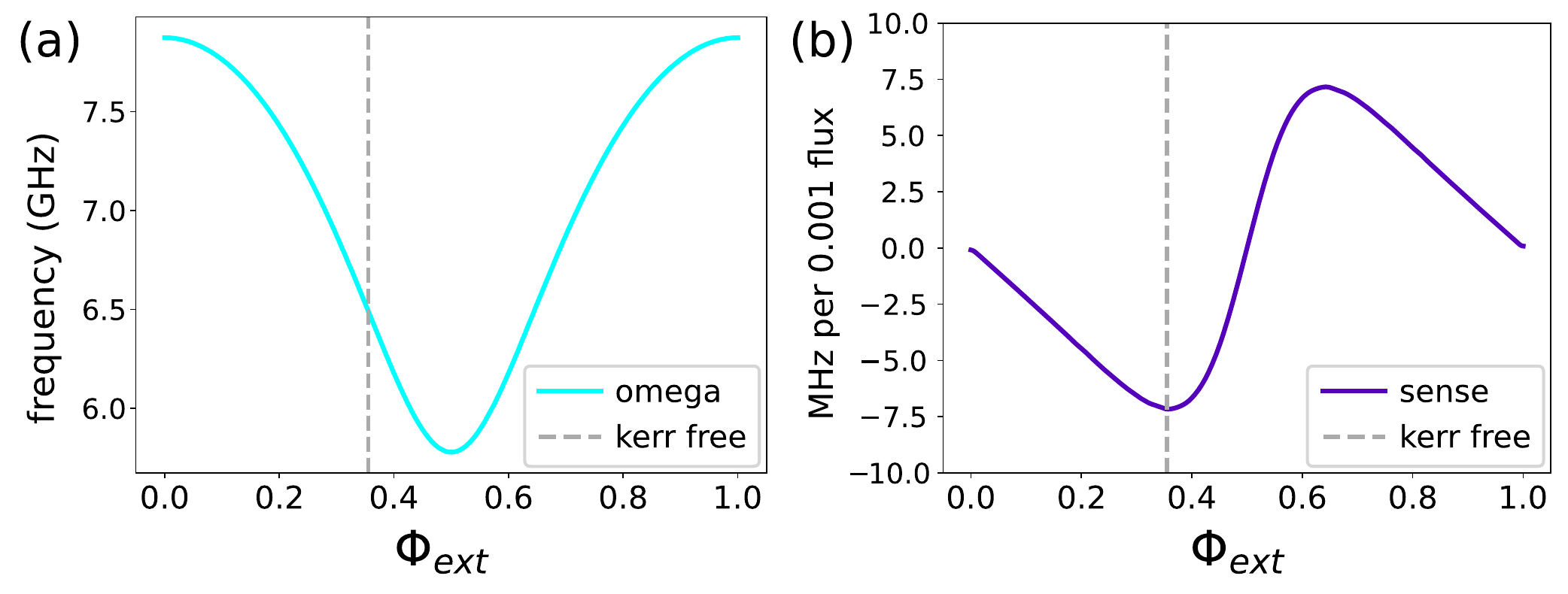}
    \caption{\textbf{SNAIL frequency dependence.} (a) SNAIL frequency and (b) frequency sensitivity, defined as $d\omega_a/d(\Phi_{\rm ext}/\Phi_0)$, versus external flux bias. The vertical dashed line indicates the KFP.}\label{snailsense}
\end{figure}

The first aspect is the ability to decouple the SNAIL and the qubit. During the trisqueezing dynamics, the SNAIL is near resonance with the HBAR, while the spectator qubit is tuned sufficiently far away so that it does not affect the evolution, despite the large SNAIL-qubit coupling of \unit{50}{MHz}. After trisqueezing we do the opposite: we tune the SNAIL sufficiently far away and bring the spectator qubit on or near resonance with the HBAR to read out the mechanical state. 
Fig.~\ref{snailsense}(a) shows the SNAIL frequency tunability. The vertical dashed line indicates the KFP at $\Phi_{\rm dc}\simeq0.399\,\Phi_0$. Note that, in the absence of a flux bias, the SNAIL frequency rises by more than \unit{1}{GHz} (leftmost point on the curve), enough to directly decouple the SNAIL from the other circuit elements.

One also needs to estimate the SNAIL-mediated coupling between the HBAR and the spectator qubit, which according to perturbation theory is
\begin{equation}
    \Delta g_{\rm qm}\approx\frac{1}{2}g_{\rm sm}g_{\rm sq}\left(\frac{1}{\omega_c-\omega_a}+\frac{1}{\omega_q-\omega_a}\right) .\label{basiccoupler}
\end{equation}
When coupling the HBAR to the spectator qubit, this SNAIL-mediated contribution adds to the direct coupling $g_{\rm qm}$. 
Setting both detunings in Eq.~\eqref{basiccoupler} to \unit{1}{GHz} gives $|\Delta g_{\rm qm}|\approx2\pi\cdot\unit{15}{kHz}$, much smaller than $g_{\rm qm}=2\pi\cdot\unit{0.24}{MHz}$, so that the SNAIL-mediated coupling only slightly modifies the qubit-HBAR interaction.

The other aspect one must take into account is frequency matching. A typical SNAIL device has one main control parameter, namely the flux bias $\Phi_{\rm ext}$. Changing this parameter affects both the mode frequency and the nonlinearity strength, and for trisqueezing to work both quantities have to stay within a suitable range, which might be hard to satisfy in realistic experimental scenarios due to fabrication errors. In fact, the SNAIL's junction sizes vary from sample to sample, which can cause the SNAIL frequency to deviate from its designed value by even a few hundred megahertz.

However, HBAR devices fortunately have a dense spectrum, with one vibration mode every free spectral range (FSR) $\sim\unit{10}{MHz}$~\cite{belles_loss_2026}, making the frequency tunability requirements much less stringent.
Figure~\ref{snailsense}(b) shows that near the KFP, the SNAIL frequency sensitivity is close to its maximum magnitude, about \unit{-7}{MHz} per $10^{-3}\,\Phi_0$.
On the other hand, our simulations show that the trisqueezing window is about $0.003\,\Phi_0$ wide, see Fig.~\ref{wignercomp}, which corresponds to a frequency tuning range of about \unit{20}{MHz}, two times larger than the HBAR FSR, so that an adjacent HBAR mode to work with can always be found.

\begin{figure}[t]
    \centering
    \includegraphics[width=0.46\textwidth]{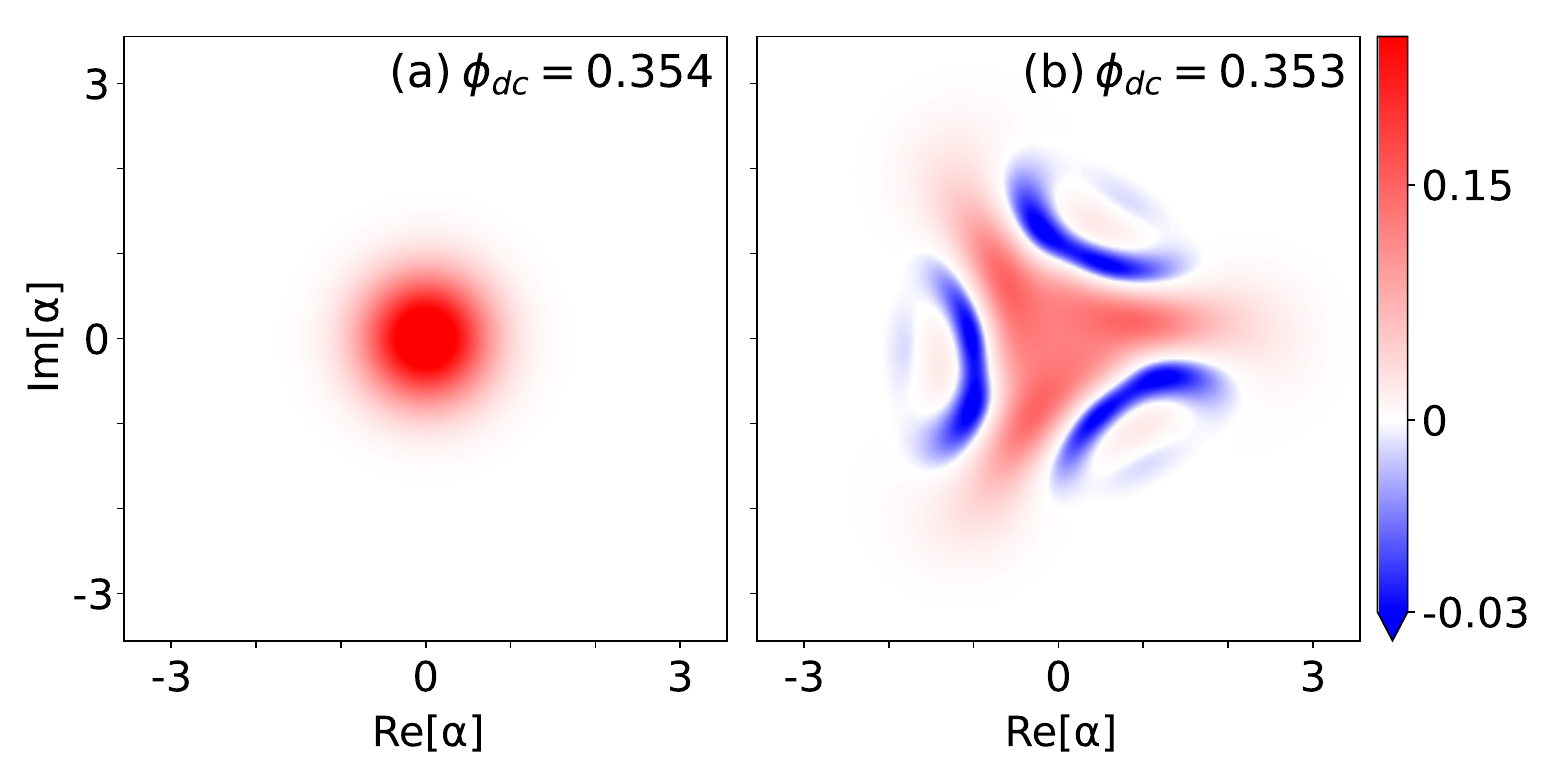}
    \caption{\textbf{Trisqueezing with compensation.} Wigner function of the HBAR state before (a) and after (b) compensating for a fabrication-induced frequency drift, at $t=0.823\,\mu\mathrm{s}$.}\label{fluxcompensation}
\end{figure}

To verify the robustness of the circuit to fabrication errors, we consider the following scenario. We take the same setting as in Fig.~\ref{wignercomp}, with the SNAIL frequency at the KFP being \unit{6.5}{GHz}, but move the nearest HBAR mode from \unit{6501.4}{MHz} to \unit{6510}{MHz}, which is equivalent to fixing the HBAR frequency while moving the SNAIL frequency. In this case, operating at the KFP no longer produces trisqueezing within a reasonable time, see Fig.~\ref{fluxcompensation}(a). To counter this issue, we can lower the flux bias by about $1.2\times10^{-3}\,\Phi_0$, which brings the SNAIL frequency to \unit{6508.2}{MHz}, only \unit{1.8}{MHz} below the HBAR mode, so that trisqueezing is restored, see Fig.~\ref{fluxcompensation}(b). 

\section{Conclusion}
In this work we have explored the generation of non-Gaussian trisqueezed states of motion in an HBAR device. In our scheme, a flux-modulated SNAIL operated at a Kerr-free point serves as the three-wave mixing element which, combined with the dispersive piezoelectric coupling to the HBAR, activates an effective three-phonon interaction in a selected mechanical mode. 
Through analytical calculations and numerical simulations of the full system Hamiltonian, we have shown that trisqueezing is achievable with realistic experimental parameters, identified viable operating points, and characterized the generated states through their skewness and quantum Fisher information. 
We have further verified the robustness of our scheme by taking into account realistic circuit parameters, the presence of a spectator readout qubit, and fabrication-induced deviations of the SNAIL frequency, which can be compensated by flux tuning thanks to the dense mode spectrum of HBARs.

Together with the Gaussian operations already realized in HBARs, namely single-mode squeezing~\cite{fadel}, beam-splitter~\cite{von_lupke_engineering_2024}, and two-mode squeezing~\cite{kern_einstein-podolsky-rosen_2026} operations, the addition of a nonlinear trisqueezing interaction completes, in principle, a gate set for universal continuous-variable quantum computing~\cite{univcv}. 
Notably, the same flux-modulated SNAIL can itself provide these Gaussian operations: through the modulated second-order term of its potential, modulating the flux at twice the frequency of a mechanical mode, or at the sum or difference of the frequencies of two modes, activates single-mode squeezing, two-mode squeezing, or beam-splitter interactions, respectively. A single nonlinear element, addressed only through the frequency of its flux modulation, could thus provide both Gaussian and non-Gaussian control of the mechanical modes. 
Combined with the long lifetimes~\cite{belles_loss_2026} and the large number of modes of HBARs, this makes the proposed platform attractive as a multimode bosonic processor and quantum memory.

Beyond quantum computing, trisqueezed states are non-Gaussian resources with negative Wigner function regions, whose enhanced sensitivity to phase-space rotations, quantified by the QFI, makes them useful for quantum metrology~\cite{fadel_quantum_2025}. In mechanical systems, phase-space rotations correspond to small shifts of the mechanical frequency, which can be induced by external perturbations of the resonator. Furthermore, since HBAR vibration modes involve macroscopic effective masses, the preparation of non-Gaussian states in these modes extends tests of quantum mechanics to increasingly massive objects~\cite{bild_schrodinger_2023}.

In the future, our platform could be extended in several directions. Since the cubic nonlinearity of the SNAIL couples to all HBAR modes, modulating the flux at the sum of three distinct mode frequencies activates the three-mode interaction $\hat c_1^\dagger\hat c_2^\dagger\hat c_3^\dagger+\hat c_1\hat c_2\hat c_3$, which generates genuine tripartite non-Gaussian entanglement~\cite{agusti_tripartite_2020} between mechanical modes. 
Moreover, a multitone flux modulation could enable the implementation of the cubic phase gate $\hat U=\exp[ir(\hat c^\dagger+\hat c)^3]$, with $r$ the cubicity, in the same system. Cubic phase states have recently been prepared in circuit QED~\cite{delsing}, and such an operation can lead to the generation of Gottesman-Kitaev-Preskill states~\cite{thegkp}, and eventually to fault-tolerant quantum computation on a continuous-variable platform. 
Finally, since the effective trisqueezing rate scales as $(g/\Delta_a)^3$, increasing the piezoelectric coupling strength, as well as optimizing the modulation through pulse shaping or optimal control, would allow for faster state preparation and higher-quality trisqueezed states.

\section*{Acknowledgments}
This work was supported by the National Natural Science Foundation of China (Grants No. U25A6009, No. 92265207, No. T2121001, No. T2322030, No. 12122504, No. 12274142, and No. 12475017), the Innovation Program for Quantum Science and Technology (Grant No. 2021ZD0301800), the Beijing Nova Program (Grant No. 20220484121), and the Ministry of Science and Technology project (Grant No. 2025YFE0217600).

M.F. was supported by the Swiss National Science Foundation Ambizione Grant No. 208886, Spark Grant No. 237484, and The Branco Weiss Fellowship -- Society in Science, administered by the ETH Z\"{u}rich.

We also thank Mei Z. Y. and Liu Y. for helpful discussions on SNAIL devices and Floquet engineering.

\clearpage
\newpage

\bibliography{sssbib}

\clearpage
\newpage

\renewcommand{\thetable}{S\arabic{table}}  
\renewcommand{\thepage}{\arabic{page}}  
\renewcommand{\thefigure}{S\arabic{figure}}
\renewcommand{\theequation}{S\arabic{equation}}
\setcounter{page}{1}
\setcounter{figure}{0}
\setcounter{table}{0}
\setcounter{section}{0}
\setcounter{equation}{0}

%\resetlinenumber
\widetext

{\centering\textbf{Supplementary Materials for:} \\} 
{\centering\textbf{Mechanical trisqueezing in circuit quantum acoustodynamics with SNAIL devices}\\} 

\normalsize
\vspace{.3cm}
{\centering Yi-Zhou Bu$^{1,2}$, Matteo Fadel$^{3,4,}$$^\ast$, Kai Xu$^{1,2}$ and Heng Fan$^{1,2,}$$^\dagger$\\
\vspace{2mm}
\textit{$^1$ Institute of Physics, Chinese Academy of Science, Beijing 100190, China\\}
\textit{$^2$ School of Physical Sciences, University of Chinese Academy of Sciences, Beijing 100049, China\\}
\textit{$^3$ Institute of Quantum Science and Technology $\&$ Frontiers Science Center for Nano-Optoelectronics, School of Physics, Peking University, Beijing 100871, China\\}
\textit{$^4$ Department of Physics, ETH Z\"{u}rich, Z\"{u}rich 8093, Switzerland\\}
\vspace{2mm}
\textit{$^\ast$ matteofadel@pku.edu.cn\\}
\textit{$^\dagger$ hfan@iphy.ac.cn}\\
}

\suppressfloats

%\tableofcontents

%\clearpage
%\newpage

\section{Detailed derivation of the SNAIL Hamiltonian}
In this section, we derive the SNAIL-mode Hamiltonian quoted in Eq.~\eqref{snailnormalhamiltonian} of the main text. The derivation proceeds by expanding the SNAIL potential around its minimum for a static flux bias, including a weak flux modulation \(\phi_{\mathrm{ac}}(t)\) to first order, adding the capacitive energy, and then quantizing the resulting weakly anharmonic mode. Finally, we show how the modulation-induced drive term is eliminated, and derive the Kerr coefficient that defines the Kerr-free point (KFP).

\subsection{SNAIL potential and expansion around a static minimum}
The SNAIL potential is
\begin{equation}
E(\phi,\phi_{\mathrm{ext}})=-\beta E_J \cos\phi-N E_J \cos\left(\frac{\phi_{\mathrm{ext}}-\phi}{N}\right)
\label{eq:SNAIL_potential_app}
\end{equation}
where \(\phi\) is the superconducting phase across the small junction, \(E_J\) is the Josephson energy of one large junction, \(N\) is the number of large junctions in the array, and \(\beta\) is the ratio between the small-junction Josephson energy and \(E_J\).
\\We consider the static case where the externally applied flux is constant $\phi_{\mathrm{ext}}=\phi_{\mathrm{dc}}$. Let \(\phi_m\equiv \phi_m(\phi_{\mathrm{dc}})\) denote the minimum of the potential at the chosen dc bias, defined implicitly by the solution
\begin{equation}
\left.
\frac{\partial E(\phi,\phi_{\mathrm{ext}})}{\partial \phi}
\right|_{\phi=\phi_m,\;\phi_{\mathrm{ext}}=\phi_{\mathrm{dc}}}
=0 \qquad\qquad \beta \sin{\phi_m} - \sin \left( \dfrac{\phi_{\rm dc}-\phi_m}{N}\right) = 0
\label{eq:def_phi_m_app}
\end{equation}
We then introduce the displacement coordinate $\varphi \equiv \phi-\phi_m$ (denoted $\delta\phi$ in the main text). Expanding the potential around \(\phi=\phi_m\), one obtains
\begin{equation}
E(\phi,\phi_{\mathrm{ext}})=E(\phi_m,\phi_{\mathrm{ext}})+\sum_{n\ge 1}\frac{1}{n!}
\left.\frac{\partial^n E(\phi,\phi_{\mathrm{ext}})}{\partial \phi^n}\right|_{\phi=\phi_m}\varphi^n\label{eq:Taylor_general_app}
\end{equation}
The linear term $\partial E/\partial\phi$ is zero for $\phi_{\mathrm{ext}}=\phi_{\mathrm{dc}}$, because \(\phi_m\) is defined to be an extremum through Eq.~\eqref{eq:def_phi_m_app}. We define the dimensionless expansion functions
\begin{equation}
E^{(n)}(\phi_{\mathrm{ext}})\equiv\left.\frac{1}{E_J}\frac{\partial^n E(\phi,\phi_{\mathrm{ext}})}{\partial \phi^n}\right|_{\phi=\phi_m}
\label{eq:def_En_app}
\end{equation}
whose values at the dc bias are the coefficients $c_n\equiv E^{(n)}(\phi_{\rm dc})$ of the main text. Equation~\eqref{eq:Taylor_general_app} then becomes (the linear term is kept for adding in flux drive later)
\begin{equation}\dfrac{E(\phi,\phi_{\mathrm{ext}})}{E_J}=E^{(0)}(\phi_{\mathrm{ext}})+E^{(1)}(\phi_{\mathrm{ext}}) \varphi
+\frac{1}{2} E^{(2)}(\phi_{\mathrm{ext}}) \varphi^2+ \cdots
\label{eq:potential_expanded_static_app}
\end{equation}
Since \(\phi_m\) is chosen to be a local minimum, one has $c_1=E^{(1)}(\phi_{\mathrm{dc}}) = 0$ and $c_2=E^{(2)}(\phi_{\mathrm{dc}}) > 0$. For completeness, the first few derivatives of Eq.~\eqref{eq:SNAIL_potential_app} with respect to \(\phi\) are
\begin{align}
\frac{\partial E}{\partial \phi}&=\beta E_J \sin\phi-E_J \sin\left(\frac{\phi_{\mathrm{ext}}-\phi}{N}\right)\\
\frac{\partial^2 E}{\partial \phi^2}&=\beta E_J \cos\phi+\frac{E_J}{N}\cos\left(\frac{\phi_{\mathrm{ext}}-\phi}{N}\right)\\
\frac{\partial^3 E}{\partial \phi^3}&=-\beta E_J \sin\phi+\frac{E_J}{N^2}\sin\left(\frac{\phi_{\mathrm{ext}}-\phi}{N}\right)\\
\frac{\partial^4 E}{\partial \phi^4}&=-\beta E_J \cos\phi-\frac{E_J}{N^3}\cos\left(\frac{\phi_{\mathrm{ext}}-\phi}{N}\right)
\end{align}
Hence
\begin{align}
E^{(1)}(\phi_{\mathrm{ext}})&=\beta \sin\phi_m-\sin\left(\frac{\phi_{\mathrm{ext}}-\phi_m}{N}\right)\label{s10}\\
E^{(2)}(\phi_{\mathrm{ext}})&=\beta \cos\phi_m+\frac{1}{N}\cos\left(\frac{\phi_{\mathrm{ext}}-\phi_m}{N}\right)\\
E^{(3)}(\phi_{\mathrm{ext}})&=-\beta \sin\phi_m+\frac{1}{N^2}\sin\left(\frac{\phi_{\mathrm{ext}}-\phi_m}{N}\right)\\
E^{(4)}(\phi_{\mathrm{ext}})&=-\beta \cos\phi_m-\frac{1}{N^3}\cos\left(\frac{\phi_{\mathrm{ext}}-\phi_m}{N}\right)
\label{eq:En_explicit_app}
\end{align}
Using Eq.~\eqref{eq:def_phi_m_app}, the static third-order coefficient can be written as $c_3=-\beta\,(1-1/N^2)\sin\phi_m$.

\subsection{First order expansion of an AC flux}
We consider the time-dependent case where the externally applied flux is $\phi_{\mathrm{ext}}(t)=\phi_{\mathrm{dc}}+\phi_{\mathrm{ac}}(t)$.
Still, we denote by \(\phi_m\equiv \phi_m(\phi_{\mathrm{dc}})\) the local minimum of the potential at the chosen dc bias $\phi_{\mathrm{dc}}$. Now, note that the linear term $E^{(1)}(\phi_{\mathrm{dc}}) = 0$, but $E^{(1)}(\phi_{\mathrm{dc}}+\phi_{\mathrm{ac}}(t)) \neq 0$. For weak flux modulations $|\phi_{\rm ac}(t)|\ll 1$ one can use a linear expansion 
\begin{equation}
    E^{(n)}(\phi_{\rm ext}) \simeq E^{(n)}(\phi_{\rm dc}) + \left.\dfrac{\partial E^{(n)}}{\partial \phi_{\rm ext}}\right\vert_{\phi_{\rm dc}}\phi_{\rm ac}(t)
\end{equation}
Combining this result with Eq.~\eqref{eq:potential_expanded_static_app} allows us to write the SNAIL potential expanded around the fixed dc operating point $\phi_m(\phi_{\rm dc})$ as
\begin{align}
    \dfrac{E(\phi,\phi_{\text{ext}})}{E_J} &= \left(E^{(0)}(\phi_{\text{dc}}) + \left.\frac{\partial E^{(0)}}{\partial\phi_{\text{ext}}}\right\vert_{\phi_{\rm dc}} \phi_{\text{ac}}(t) \right) \notag \\
    &\quad + \left(\left.\frac{\partial E^{(1)}} {\partial\phi_{\text{ext}}}\right\vert_{\phi_{\rm dc}} \phi_{\text{ac}}(t)\right) \varphi \notag\\
    &\quad + \frac{1}{2}\left(E^{(2)}(\phi_{\text{dc}}) + \left.\frac{\partial E^{(2)}}{\partial\phi_{\text{ext}}}\right\vert_{\phi_{\rm dc}} \phi_{\text{ac}}(t)\right)\varphi^2 \nonumber\\
    &\quad + \frac{1}{6}\left(E^{(3)}(\phi_{\text{dc}}) + \left.\frac{\partial E^{(3)}}{\partial\phi_{\text{ext}}}\right\vert_{\phi_{\rm dc}} \phi_{\text{ac}}(t)\right)\varphi^3 \nonumber\\
    &\quad + \frac{1}{24}\left(E^{(4)}(\phi_{\text{dc}}) + \left.\frac{\partial E^{(4)}}{\partial\phi_{\text{ext}}}\right\vert_{\phi_{\rm dc}} \phi_{\text{ac}}(t)\right)\varphi^4
    \label{appendix:snailexpand}
\end{align}
As in the main text, we define the modulated coefficients $c_n'\equiv\partial E^{(n)}/\partial\phi_{\rm ext}|_{\phi_{\rm dc}}$, where the derivative is taken at fixed $\phi$. Note that the term in the second line is in general nonzero. Explicitly, from Eqs.~\eqref{s10}--\eqref{eq:En_explicit_app},
\begin{equation}
    c_1' = - \dfrac{1}{N} \cos\theta_m,\qquad c_2'=-\frac{1}{N^2}\sin\theta_m,\qquad c_3'=\frac{1}{N^3}\cos\theta_m,\qquad c_4'=\frac{1}{N^4}\sin\theta_m,\qquad \theta_m\equiv\frac{\phi_{\rm dc}-\phi_m}{N} .
\end{equation}
With $c_n$ and $c_n'$, Eq.~\eqref{appendix:snailexpand} coincides with Eq.~\eqref{snailexpand} of the main text.

\subsection{Addition of the capacitive energy and quantization}\label{appendix:quant}
To obtain the Hamiltonian of the dynamical mode, the potential energy must be supplemented by the capacitive energy. Denoting by \(Q\) the charge canonically conjugate to \(\phi\), the electrostatic energy is
\begin{equation}
T = \frac{Q^2}{2C}\label{eq:cap_energy_Q_app}
\end{equation}
where \(C\) is the effective capacitance of the mode. Introducing the Cooper-pair number $n = Q/2e$ and the charging energy $E_C \equiv e^2/(2C)$, the capacitive contribution becomes
\begin{equation}
T = 4E_C n^2\label{eq:cap_energy_n_app}
\end{equation}
Upon quantization, \(\varphi\) and \(n\) become operators obeying $[\hat\varphi,\hat n]=i$, the full Hamiltonian is then
\begin{equation}
\hat{H}=4E_C \hat{n}^2+E_J \sum_{k=0}^{\infty} \frac{1}{k!}\left[c_k+c_k'\, \phi_{\mathrm{ac}}(t)\right]\hat{\varphi}^{\,k}\label{apeq:fullham}
\end{equation}
We express $\hat{\varphi}$ and $\hat{n}$ through ladder operators as $\hat{\varphi} = \varphi_{\mathrm{zpf}}(\hat{a}_0 + \hat{a}_0^\dagger)$, $\hat{n} = i\,n_{\mathrm{zpf}}(\hat{a}_0^\dagger - \hat{a}_0)$, with zero-point fluctuations $\varphi_{\mathrm{zpf}}$, $n_{\mathrm{zpf}}$ and mode frequency $\omega_a^0$ given by ($\hbar=1$)
\begin{equation}
\varphi_{\mathrm{zpf}}=\left(\frac{2E_C}{E_J c_2}\right)^{1/4}
\qquad
2\varphi_{\mathrm{zpf}} n_{\mathrm{zpf}} = 1
\qquad
\omega_a^0=\sqrt{8E_C E_J c_2}
\end{equation}
If the static quadratic sector is diagonalized exactly, then the Hamiltonian can be written as
\begin{equation}
\hat{H}=\omega_a^0\left(\hat{a}_0^\dagger \hat{a}_0 + \frac{1}{2}\right)+\sum_{n \geq 0} g^{(n)}(t)\,(\hat{a}_0 + \hat{a}_0^\dagger)^n
\end{equation}
up to an irrelevant constant shift already contained in \(g^{(0)}(t)\), the coefficients are
\begin{equation}
g^{(n)}(t)=\frac{E_J}{n!}\,E^{(n)}\!\bigl(\phi_{\mathrm{ext}}(t)\bigr)\,\varphi_{\mathrm{zpf}}^{\,n}
=\frac{E_J}{n!}\left[c_n+c_n'\,\phi_{\mathrm{ac}}(t)\right]\varphi_{\mathrm{zpf}}^{\,n}
\end{equation}
We can separate $g^{(n)}(t)$ into static and modulated parts $g^{(n)}(t)\equiv g^{(n)}_{\rm dc}+g^{(n)}_{\rm ac}\phi_{\mathrm{ac}}(t)$
\begin{equation}
g^{(n)}_{\rm dc}=\frac{1}{n!}E_J\,c_n\,\varphi_{\mathrm{zpf}}^n
\qquad
g^{(n)}_{\rm ac}=\frac{1}{n!}E_J\,c_n'\,\varphi_{\mathrm{zpf}}^n\label{definition:gn}
\end{equation}
This notation is consistent with Eq.~\eqref{snailcouplings} of the main text, and the values used in the numerical simulations are listed in Table~\ref{snailtable}, where the entry $g^{(2)}_{\rm dc}$ refers to the static quadratic coefficient $E_Jc_2\varphi_{\rm zpf}^2/2=\omega_a^0/4$ before it is absorbed into $\omega_a^0$. Because the static quadratic part has been absorbed into $\omega_a^0$ (so that $g^{(2)}_{\rm dc}\equiv0$ in the main text) while $g^{(1)}_{\rm dc}$ is just 0, we can express the full Hamiltonian as
\begin{equation}
\hat{H}=\omega_a^0\left(\hat{a}_0^\dagger \hat{a}_0 + \frac{1}{2}\right)+ g^{(1)}_{\rm ac}\phi_{\mathrm{ac}}(t)(\hat{a}_0 + \hat{a}_0^\dagger)+ g^{(2)}_{\rm ac}\phi_{\mathrm{ac}}(t)(\hat{a}_0 + \hat{a}_0^\dagger)^2+ g^{(3)}(t)(\hat{a}_0 + \hat{a}_0^\dagger)^3+ g^{(4)}(t)(\hat{a}_0 + \hat{a}_0^\dagger)^4+\cdots\label{gntexpand}
\end{equation}
Truncating at fourth order and omitting the zero-point energy, Eq.~\eqref{gntexpand} is just Eq.~\eqref{snailnormalhamiltonian} in main text. Note the term $g^{(2)}_{\rm ac}\phi_{\rm ac}(t)(\hat{a}_0+\hat{a}_0^\dagger)^2$ contains both frequency modulation and squeezing, expand it as $(\hat{a}_0+\hat{a}_0^\dagger)^2=\hat{a}_0^2+(\hat{a}_0^\dagger)^2+2\hat{a}_0^\dagger \hat{a}_0+1$, neglect the constant 1, so that terms associated with 2nd order are
\begin{equation}
\hat{H}_2=\omega_a^0\left(\hat{a}_0^\dagger \hat{a}_0+\frac{1}{2}\right)+2g^{(2)}_{\rm ac}\phi_{\rm ac}(t)\hat{a}_0^\dagger \hat{a}_0+g^{(2)}_{\rm ac}\phi_{\rm ac}(t)\bigl(\hat{a}_0^2+(\hat{a}_0^\dagger)^2\bigr)
\end{equation}
We obtain an instantaneous frequency modulation $\omega_a(t)=\omega_a^0+2g^{(2)}_{\rm ac}\phi_{\rm ac}(t)$ up to a \(c\)-number shift. At the same time, the terms $g^{(2)}_{\rm ac}\phi_{\rm ac}(t)\bigl(\hat{a}_0^2+(\hat{a}_0^\dagger)^2\bigr)$ are the single-mode squeezing terms, which become resonant for modulation frequencies close to $2\omega_a^0$.

\subsection{Elimination of the modulation-induced drive}\label{appendix:disp}
The term $g^{(1)}_{\rm ac}\phi_{\rm ac}(t)(\hat a_0+\hat a_0^\dagger)$ in Eq.~\eqref{gntexpand} is a drive on the SNAIL mode. It accounts for the motion of the potential minimum induced by the flux modulation, which is not captured by the static expansion point $\phi_m(\phi_{\rm dc})$. We remove it with the time-dependent displacement $\hat a_0\rightarrow\hat a_0+\alpha(t)$, corresponding to $\hat\varphi\rightarrow\hat\varphi+\delta\phi_{\rm cl}(t)$ with $\delta\phi_{\rm cl}=\varphi_{\rm zpf}[\alpha(t)+\alpha^*(t)]$. Including the term $-i\hat D^\dagger\dot{\hat D}$ generated by the time-dependent displacement operator $\hat D$, the terms linear in $\hat a_0$ originating from the harmonic part and from the drive cancel provided that
\begin{equation}
i\dot\alpha=\omega_a^0\,\alpha+g^{(1)}_{\rm ac}\phi_{\rm ac}(t) .
\end{equation}
For a harmonic modulation $\phi_{\rm ac}(t)=\phi_{\rm ac}\cos(\omega_pt)$, the steady-state solution is $\alpha(t)=-\frac{1}{2}g^{(1)}_{\rm ac}\phi_{\rm ac}\left[e^{-i\omega_pt}/(\omega_a^0-\omega_p)+e^{i\omega_pt}/(\omega_a^0+\omega_p)\right]$. Using $g^{(1)}_{\rm ac}=E_Jc_1'\varphi_{\rm zpf}$, $\omega_a^0\varphi_{\rm zpf}^2=4E_C$ and $(\omega_a^0)^2=8E_CE_Jc_2$, one finds
\begin{equation}
\delta\phi_{\rm cl}(t)=-\frac{2\omega_a^0\,g^{(1)}_{\rm ac}\varphi_{\rm zpf}}{(\omega_a^0)^2-\omega_p^2}\,\phi_{\rm ac}(t)=\chi\,\phi_{\rm ac}(t),\qquad
\chi=-\frac{c_1'}{c_2}\,\frac{(\omega_a^0)^2}{(\omega_a^0)^2-\omega_p^2} ,
\end{equation}
as stated in the main text. Substituting $\hat\varphi\rightarrow\hat\varphi+\chi\phi_{\rm ac}(t)$ into the static nonlinear terms $E_Jc_n\hat\varphi^n/n!$ and keeping terms to first order in $\phi_{\rm ac}$ generates $E_J\chi c_n\phi_{\rm ac}(t)\hat\varphi^{n-1}/(n-1)!$, which adds to the modulated term of order $n-1$, while the corresponding corrections to the modulated terms are of order $\phi_{\rm ac}^2$. The modulated coefficients are thus renormalized as
\begin{equation}
c_n'\rightarrow\tilde c_n'=c_n'+\chi\,c_{n+1},\qquad g^{(n)}_{\rm ac}\rightarrow\tilde g^{(n)}_{\rm ac}=\frac{1}{n!}E_J\,\tilde c_n'\,\varphi_{\rm zpf}^n .
\end{equation}
For $\omega_p\ll\omega_a^0$ the displacement follows the minimum adiabatically, $\chi\rightarrow-c_1'/c_2=d\phi_m/d\phi_{\rm dc}$ (as obtained by differentiating $E^{(1)}=0$), so that $\tilde c_n'\rightarrow dc_n/d\phi_{\rm dc}$. For the KFP considered in the main text one finds $\chi\simeq1.24$ and $\tilde c_3'\simeq0.072$ in this limit, while for $\omega_p\simeq3\omega_a^0$, relevant for trisqueezing, $\chi\simeq-0.16$ and $\tilde c_3'\simeq0.033$. Modulation frequencies $\omega_p\simeq\omega_a^0$, for which $\chi$ diverges, must be avoided.

\subsection{Kerr coefficient and Kerr-free point}\label{appendix:kerr}
In the absence of modulation, the static terms $g^{(3)}_{\rm dc}(\hat a_0+\hat a_0^\dagger)^3$ and $g^{(4)}_{\rm dc}(\hat a_0+\hat a_0^\dagger)^4$ shift the levels $|N\rangle$ of the SNAIL mode. To first order in $g^{(4)}_{\rm dc}$ and second order in $g^{(3)}_{\rm dc}$ (see Eq.~\eqref{eq:floquete0} below),
\begin{equation}
E_N=N\omega_a^0+g^{(4)}_{\rm dc}\left(6N^2+6N+3\right)-\frac{\left(g^{(3)}_{\rm dc}\right)^2}{\omega_a^0}\left(30N^2+30N+11\right) .
\end{equation}
Writing $N^2+N=N(N-1)+2N$, the $N$-dependent part takes the form $(K/2)N(N-1)+\delta\omega\,N$, i.e., it corresponds to the effective Hamiltonian $(\omega_a^0+\delta\omega)\hat a_0^\dagger\hat a_0+(K/2)\hat a_0^{\dagger2}\hat a_0^2$, with
\begin{equation}
K=\delta\omega=12g^{(4)}_{\rm dc}-\frac{60\left(g^{(3)}_{\rm dc}\right)^2}{\omega_a^0}=\frac{E_C}{c_2}\left(c_4-\frac{5c_3^2}{3c_2}\right) .
\end{equation}
The KFP is defined by $K=0$, i.e., $c_4=5c_3^2/(3c_2)$, where also the static frequency shift $\delta\omega$ vanishes at this order, so that $\omega_a^0$ is the SNAIL frequency. For $N=3$ and $\beta=0.1$, the KFP is at $\Phi_{\rm dc}\simeq0.399\,\Phi_0$ ($\phi_{\rm dc}\simeq2.51$, $\phi_m\simeq2.28$), where $c_2\simeq0.267$, $c_3\simeq-0.067$, $c_4\simeq0.028$ and $c_3'\simeq0.037$.

\clearpage
\newpage

\section{Derivation of the effective trisqueezing Hamiltonian}\label{appendix:trisq}

Starting from a two-mode Hamiltonian where (i) a linear beam-splitter exchange couples modes $a$ and $c$ and (ii) a parametrically driven cubic nonlinearity acts on mode $a$, we derive an effective Hamiltonian for mode $c$ alone. In the dispersive limit $|\Delta_a|\gg g$ we want to show that the parametric drive generates an induced cubic interaction on $c$,
\[
\hat H_{\mathrm{cubic}}(t)\ = -\Big(\frac{g}{\Delta_a}\Big)^3 \epsilon(t)\,(\hat c^\dagger+\hat c)^3,
\]
and then we show how a suitable choice of parametric drive frequencies makes the desired trisqueeze interaction resonant in the rotating frame of mode $c$.

\subsection{Cubic interaction through mode dressing}
We take two bosonic modes $a,c$ with bare frequencies $\omega_a^0,\omega_c$ and a beam-splitter coupling $g$:
\begin{align}
\hat H(t) &= \hat H_0 + \hat V + \hat H_3(t),
\\
\hat H_0 &= \omega_a^0 \hat a^\dagger \hat a + \omega_c \hat c^\dagger \hat c,
\\
\hat V &= g\,(\hat a^\dagger \hat c + \hat a \hat c^\dagger),
\end{align}
then assume a parametrically driven cubic nonlinearity on the $a$ mode, namely
\begin{equation}
\hat H_3(t) = \epsilon(t)\,(\hat a+\hat a^\dagger)^3,\label{eq:adrive}
\end{equation}
where $\epsilon(t)=\tilde g^{(3)}_{\rm ac}\phi_{\rm ac}(t)$ is the modulated cubic coefficient of the SNAIL, renormalized as in Appendix~\ref{appendix:disp}. As in the main text, the static nonlinear terms and the other modulated terms are non-resonant and are omitted here. Here $\hat a$ and $\hat c$ denote the bare SNAIL and HBAR modes, written $\hat a_0$ and $\hat c_0$ in the main text.
% To reproduce a ``full cubic'' interaction in the rotating frame (i.e.\ to keep simultaneously the terms rotating at $\omega_c$ and $3\omega_c$), we will allow a two-tone modulation of the cubic amplitude:
Define the detuning $\Delta_a \equiv \omega_a^0-\omega_c$ and work in the dispersive regime
$|\Delta_a|\gg g$, so that mode $a$ can be perturbatively eliminated via a Schrieffer--Wolff transformation. Throughout we keep terms to the lowest nontrivial order required to obtain third order interaction on mode $c$.

\bigskip

The Schrieffer--Wolff elimination of the beam-splitter interaction is achieved by choosing a suitable SW generator, we seek a unitary transformation
\begin{equation}
\hat H' = e^{\hat S} \hat H e^{-\hat S},\qquad \hat S^\dagger=-\hat S,\label{twequation}
\end{equation}
that removes $\hat V$ to first order in $g/\Delta_a$. It is common to choose
\begin{equation}
    \lambda \equiv \frac{g}{\Delta_a},
    \qquad
    \hat S = \lambda(\hat a^\dagger \hat c - \hat a \hat c^\dagger),
    \label{eq:S_choice}
\end{equation}
so that $[\hat S,\hat H_0]=-\hat V$. Expand \eqref{twequation} to get the effective time-independent Hamiltonian to second order
\begin{equation}
\hat H' = e^{\hat S}(\hat H_0+\hat V)e^{-\hat S}
= \hat H_0 + \hat V + [\hat S,\hat H_0+\hat V] + \frac{1}{2}[\hat S,[\hat S,\hat H_0+\hat V]] + \cdots,
\end{equation}
Using $[\hat S,\hat H_0]=-\hat V$ to cancel $\hat V$ to first order and evaluate the rest of the commutators, we obtain
\begin{align}
\hat H' &= \hat H_0 + \frac{1}{2}[\hat S,\hat V] + O(\lambda^3)\\
&= \Big(\omega_a^0+\frac{g^2}{\Delta_a}\Big)\hat a^\dagger \hat a
+ \Big(\omega_c-\frac{g^2}{\Delta_a}\Big)\hat c^\dagger \hat c
+ O\!\left(\frac{g^3}{\Delta_a^2}\right).
\label{eq:Hprime_basic}
\end{align}
The terms in parentheses are dispersive shifted frequencies, substitute them as $\omega_a'=\omega_a^0+g^2/\Delta_a$, $\omega_c'=\omega_c-g^2/\Delta_a$, where $g^2/\Delta_a$ is the Lamb shift of the phonon-like mode discussed in the main text. The crucial step for trisqueezing lies in the SW transformation of $\hat H_3(t)$:
\begin{equation}
\hat H_3'(t) \equiv e^{\hat S} \hat H_3(t) e^{-\hat S}
\simeq \epsilon(t)\,(\hat a'+\hat a'^\dagger)^3 \;.\label{h3tsw}
\end{equation}
$\hat a'$ is the SW transformed operator $\hat a$:
\begin{align}
\hat a' \equiv e^{\hat S} \hat a e^{-\hat S}
&= \hat a + [\hat S,\hat a] + \frac{1}{2}[\hat S,[\hat S,\hat a]] + O(\lambda^3)\\
&= \hat a - \lambda \hat c -\frac{\lambda^2}{2}\hat a + O(\lambda^3),
\end{align}
Since $\hat a'$ contains $\hat c$, this effect is usually referred to as mode dressing. 
To leading order in $\lambda=g/\Delta_a$ we keep only $\hat a' \approx \hat a - \lambda \hat c$, which is the hybridization $\hat a_0\simeq\hat a-(g/\Delta_a)\,\hat c$ of the main text, with $\hat a$ and $\hat c$ now denoting the SNAIL-like and phonon-like dressed modes. Define $\hat X\equiv \hat a+\hat a^\dagger$ and $\hat Y\equiv \hat c+\hat c^\dagger$, so that \eqref{h3tsw} can be expanded as:
\begin{align}
\hat H_3'(t)&\approx
\epsilon(t)(\hat a'+\hat a'^\dagger)^3 \approx \epsilon(t)(\hat X-\lambda \hat Y)^3\nonumber\\
&= \epsilon(t)\hat X^3 - 3\lambda\epsilon(t) \hat X^2 \hat Y + 3\lambda^2\epsilon(t) \hat X \hat Y^2 - \lambda^3\epsilon(t) \hat Y^3.
\label{eq:cubic_expand}
\end{align}
The last term is the cubic interaction demanded by a trisqueeze gate, acting purely on mode $c$:
\begin{equation}
\hat H_{\mathrm{cubic}}(t)
=
-\lambda^3\epsilon(t)\,(\hat c+\hat c^\dagger)^3
=
-\left(\frac{g}{\Delta_a}\right)^3 \epsilon(t)\,(\hat c+\hat c^\dagger)^3.
\label{eq:induced_cubic_time}
\end{equation}
In \eqref{eq:cubic_expand}, the terms proportional to $\hat X^3$, $\hat X^2\hat Y$, $\hat X\hat Y^2$ involve mode $a$, these will be the main limiting factors on overall process fidelity and their effects cannot be simply overlooked. For a typical elimination scheme we assume
\begin{itemize}
    \item (i) mode $a$ remains near vacuum (or a small manifold) and/or is strongly damped, 
    \item (ii) these couplings are off-resonant in the chosen rotating frame.
\end{itemize}
For the trisqueezing drive discussed below, the least detuned of these terms are $\hat a^\dagger \hat c^{\dagger2}$ and $\hat a^{\dagger2}\hat c^\dagger$ (together with their Hermitian conjugates), detuned by $\simeq\Delta_a$ and $2\Delta_a$, respectively.

\subsection{Trisqueezing and cubic phase gate}%Two-tone choice and rotating-wave approximation (RWA)
Move to the rotating frame of $\omega_c'$ so that $\hat c(t)=\hat c\,e^{-i\omega_c' t}$. In this rotating frame $\hat H_{\mathrm{cubic}}(t)$ contains terms rotating at either $\omega_c'$ or $3\omega_c'$:
\begin{equation}
(\hat c+\hat c^\dagger)^3
\rightarrow\big(\hat c e^{-i\omega_c' t}+\hat c^\dagger e^{+i\omega_c' t}\big)^3
=\hat c^3 e^{-i3\omega_c' t}+3\hat c\hat c^\dagger \hat ce^{-i\omega_c' t}+3\hat c^\dagger \hat c\hat c^\dagger e^{+i\omega_c' t}+\hat c^{\dagger 3} e^{+i3\omega_c' t}.
\label{eq:cubic_rotating_components}
\end{equation}
To preserve these third order interactions, we will allow a two-tone modulation:
\begin{equation}
\epsilon(t)=\Omega_{1}\cos(\omega_{d1}t+\phi_1)+\Omega_{3}\cos(\omega_{d3}t+\phi_3),
\label{eq:two_tone_eps}
\end{equation}
where $\Omega_1$ and $\Omega_3$ are the cubic amplitudes of the two tones. Since the renormalization factor $\chi$ of Appendix~\ref{appendix:disp} depends on the modulation frequency, $\tilde c_3'$ has to be evaluated separately for each tone. Tone $\omega_{d1}$ is set to be near $\omega_c'$, $\omega_{d3}$ near $3\omega_c'$. From now on we omit the primes in $\omega_c'$ for convenience and define the drive detunings
\begin{equation}
\delta_1=\omega_{d1}-\omega_c,\qquad
\delta_3=\omega_{d3}-3\omega_c,
\end{equation}
Inserting \eqref{eq:cubic_rotating_components} and \eqref{eq:two_tone_eps} into \eqref{eq:induced_cubic_time} and expanding the cosines, we obtain
\begin{align}
\hat H_{\mathrm{cubic}}(t)
=
-\frac{\lambda^3}{2}
&\Big(
\Omega_1 e^{+i(\omega_{d1}t+\phi_1)}+\Omega_3 e^{+i(\omega_{d3}t+\phi_3)}+h.c.
\Big)\nonumber
\\
\times&\Big(
\hat c^3 e^{-i3\omega_c t}
+3\hat c\hat c^\dagger \hat c e^{-i\omega_c t}
+3\hat c^\dagger \hat c\hat c^\dagger e^{+i\omega_c t}
+\hat c^{\dagger 3} e^{+i3\omega_c t}
\Big).\label{hcubict}
\end{align}

\paragraph{Trisqueeze gate}
Drive with single tone $\omega_{d3}$ so that $\Omega_1=0$, under this situation $\hat c^3$ and $\hat c^{\dagger3}$ are resonant, all other pairings are dropped under RWA, thus:
\begin{equation}
\hat H_{\rm tsq}\approx-\frac{\lambda^3\Omega_3}{2}\Big(e^{i(\delta_3 t+\phi_3)}\hat c^3+e^{-i(\delta_3 t+\phi_3)}\hat c^{\dagger 3}\Big)
\xrightarrow{\delta_3=\phi_3=0}
-\frac{\lambda^3\Omega_3}{2}(\hat c^{\dagger3}+\hat c^3),
\label{eq:H_3omega}
\end{equation}
This is the trisqueeze interaction we put our main focus on. With $\omega_{d3}=3\omega_c'=3(\omega_c-g^2/\Delta_a)$ and $\Omega_3=\tilde g^{(3)}_{\rm ac}\phi_{\rm ac}$, it coincides with Eq.~\eqref{trisqsnailhbarham} of the main text.
\bigskip

\paragraph{Combined effective resonant cubic interaction and cubic phase gate}
The contribution from $\Omega_1 \cos(\omega_{d1}t+\phi_1)$ makes mixed terms $\hat c^\dagger \hat c\hat c^\dagger$ and $\hat c\hat c^\dagger \hat c$ resonant. When both drives $\omega_{d1}$ and $\omega_{d3}$ are present, let them share the same amplitude and phase $\phi_1=\phi_3\equiv \phi$, $\Omega_1=\Omega_3\equiv \Omega$, under exact resonance ($\delta_1=\delta_3=0$) condition all four terms in \eqref{hcubict} are resonant:
\begin{equation}
\hat H_{\mathrm{cubic}}(t)
\approx
-\frac{\lambda^3\Omega}{2}
\left(
e^{i\phi}\big[\hat c^3+3\hat c\hat c^\dagger \hat c\big]
+
e^{-i\phi}\big[\hat c^{\dagger 3}+3\hat c^\dagger \hat c\hat c^\dagger\big]
\right).
\end{equation}
Setting $\phi=0$ (drive in phase), we obtain a full cubic interaction on mode $c$, with strength suppressed by $(g/\Delta_a)^3$:
\begin{equation}
\hat H_{\mathrm{cup}}
\approx
-\frac{\lambda^3\Omega}{2}
\left(
\hat c^3 + 3\hat c\hat c^\dagger \hat c + 3\hat c^\dagger \hat c \hat c^\dagger + \hat c^{\dagger 3}
\right)
=
-\frac{\lambda^3\Omega}{2}\,(\hat c+\hat c^\dagger)^3.
\label{eq:final_full_cubic}
\end{equation}
Note that the tone $\omega_{d1}\simeq\omega_c$ lies only $|\Delta_a|$ away from the SNAIL frequency, where the displacement factor $\chi$ of Appendix~\ref{appendix:disp} is resonantly enhanced, so that this tone also strongly drives the SNAIL mode (see Appendix~\ref{appendix:xyy}).
If $\hat H_{\mathrm{cup}} = \kappa (\hat c+\hat c^\dagger)^3$ is applied for time $T$, the unitary time evolution operator is:
\begin{equation}
\hat U(T)=\exp\!\big(-i \kappa T (\hat c+\hat c^\dagger)^3\big)
=
\exp\!\big(-i \gamma\,\hat x^3\big),
\quad
\hat x=\frac{\hat c+\hat c^\dagger}{\sqrt{2}},
\quad
\gamma = \kappa T ( \sqrt{2})^3 = 2\sqrt{2}\,\kappa T.
\end{equation}
This is the form of a cubic phase gate on mode $c$, $\hat U=e^{ir(\hat c^\dagger+\hat c)^3}$ with cubicity $r=-\kappa T$ (see main text).

\clearpage
\newpage

\section{Challenges on realizing a real cubic phase gate}\label{appendix:xyy}
As mentioned earlier, cross-terms between modes $a$ and $c$ in $\hat H_3'(t)$ Eq.~\eqref{eq:cubic_expand}  can significantly affect the gates targeted on mode $c$. 
Here we investigate their real impact. 
We first study the case of trisqueezing, to then look at the cubic phase gate. 
We simulate a system evolving through $\hat H_3'(t)$ under trisqueezing conditions, with/without the $\hat X^2\hat Y$ and $\hat X\hat Y^2$ terms. 
Because the interactions we are concerned with are all to third order, unlike the simulations in main text we neglect here other SNAIL nonlinearities.
\begin{figure}[h]
    \centering
    \subfloat[keep all terms in $\hat H_3'(t)$]{\includegraphics[width=0.26\textwidth]{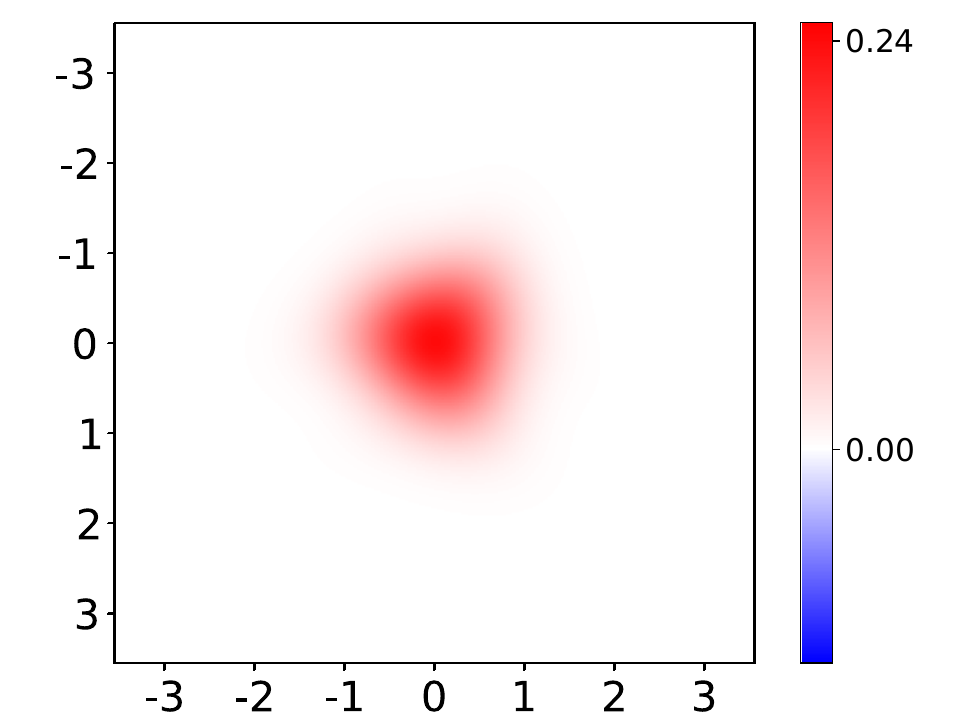}\label{apfig-wxyy}}
    \hskip 5pt
    \subfloat[drop $\hat X^2\hat Y$ and $\hat X\hat Y^2$ in $\hat H_3'(t)$]{\includegraphics[width=0.26\textwidth]{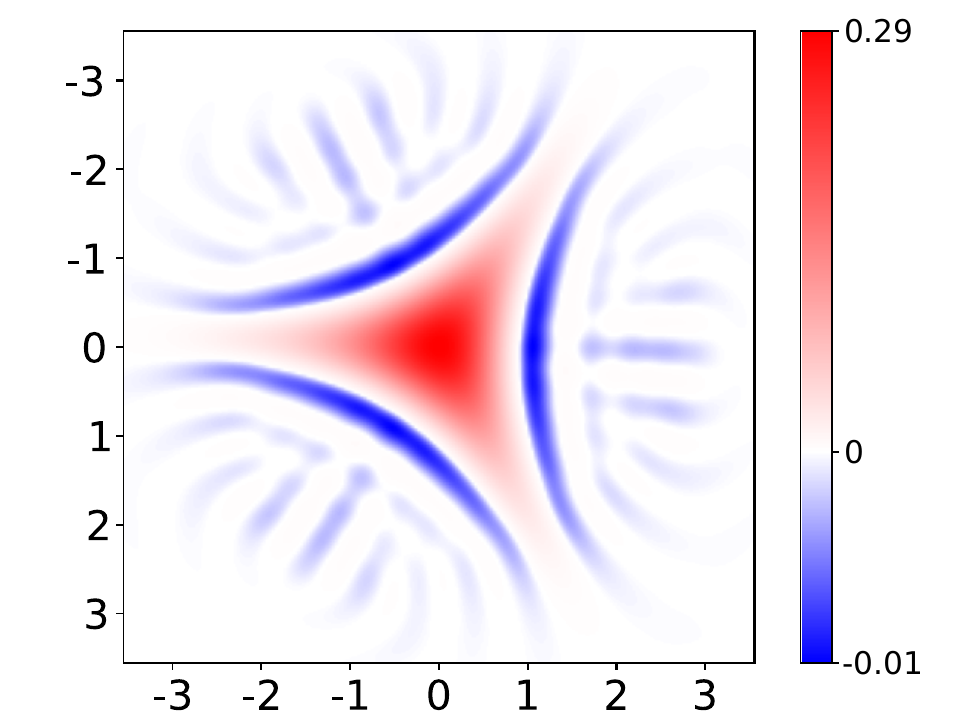}\label{apfig-nxyy}}
    \hskip 5pt
    \subfloat[keep all terms in $\hat H_3'(t)$, while adding huge damping to SNAIL mode, damping rate $\kappa_a=10\,\mu\mathrm{s}^{-1}$]{\includegraphics[width=0.26\textwidth]{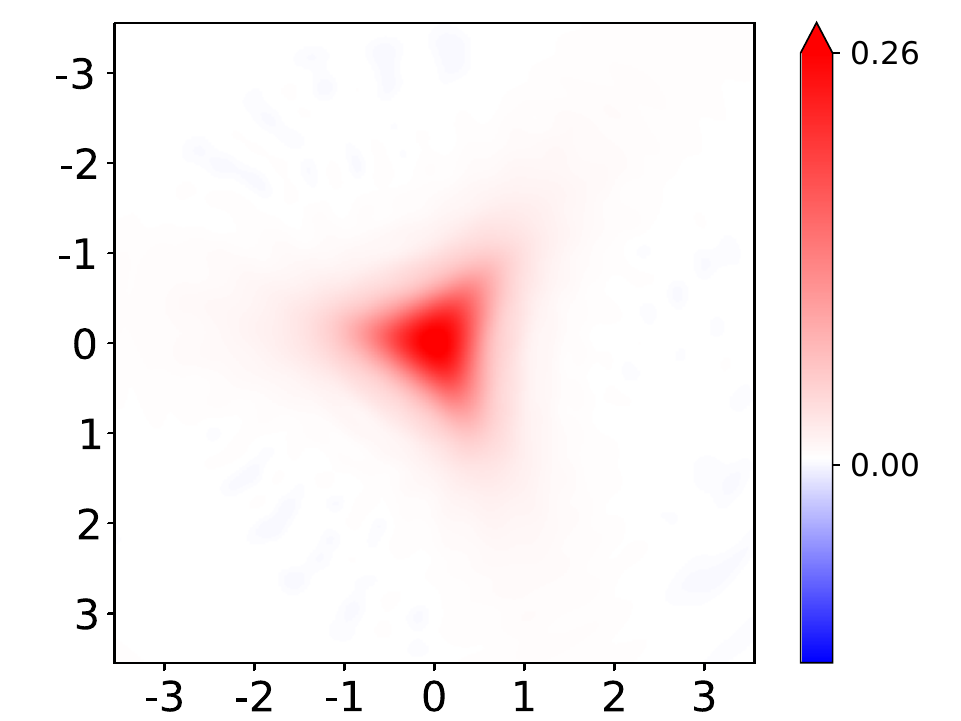}\label{apfig-damp}}
    \caption{\label{apfig-wig}Evolution of $|0\rangle$ under different Hamiltonians, trisqueeze drive $\epsilon(t)=\Omega_3\cos(\omega_{d3}t)$ with $\omega_{d3}=3\omega_c$, $T=0.5\,\mu\mathrm{s}$.}
\end{figure}
\\Fig.~\ref{apfig-wig} shows the outcome of simulations, the middle one Fig.~\ref{apfig-nxyy} discards all cross terms and closely resembles an ideal trisqueeze state, similarly to Fig.~\ref{trisqideal} of the main text. 
Fig.~\ref{apfig-wxyy} uses all terms in $\hat H_3'(t)$ and exhibits a largely degraded state fidelity. 
One way to partially prevent this is to constrain SNAIL excitations, for instance, by adding significant damping $\kappa_a$ to it so it stays mostly in ground state, such that the effect of terms like $\hat a^{\dagger2}\hat c$ would be less prominent. 
Also, one should not overdo the damping, as it will result in a significant Purcell effect~\cite{sete} causing the HBAR mode energy to leak out through SNAIL at rate
\begin{equation}
    \Gamma_{\mathrm{P}}=\frac{g^2}{\Delta_a^2}\kappa_a  . \label{eqpurcell}
\end{equation}
In our case, the detuning $\Delta_a$ is not too large compared to $g$, which will make $\Gamma_{\mathrm{P}}$ much larger than HBAR's intrinsic damping rate $\kappa_i$. 
For the sake of balancing, we keep $g/\Delta_a=0.25$ (consistent with main text), $\kappa_a=10\,\mu\mathrm{s}^{-1}$, $t=0.5\,\mu\mathrm{s}$. 
This ensures that $\Gamma_{\mathrm{P}}t\approx 0.3<1$. Because of the large damping, we can truncate the SNAIL's Hilbert space to only 2 levels (normally we use 30 for dissipation-less systems). 
The result is shown in Fig.~\ref{apfig-damp}. 
Despite discrepancies, it is still an improvement compared to Fig.~\ref{apfig-wxyy}. After rotating the state by some angle to align it with Fig.~\ref{apfig-nxyy} we find a 0.83 fidelity.

The above observations confirm that, although $\hat X^2\hat Y$ and $\hat X\hat Y^2$ are off-resonant terms, their impact is still prominent. For trisqueezing, their least detuned components, $\hat a^\dagger \hat c^{\dagger2}$ and $\hat a^{\dagger2}\hat c^\dagger$, are detuned by only $\simeq\Delta_a$ and $2\Delta_a$. When the 1x drive tone $\omega_{d1}$ is applied, some of their components even become resonant and, as we are going to show below, this can be detrimental to the realization of cubic phase gate.

\begin{figure}[h]
    \centering
    \subfloat[drop $\hat X^2\hat Y$ and $\hat X\hat Y^2$ in $\hat H_3'(t)$]{\includegraphics[width=0.26\textwidth]{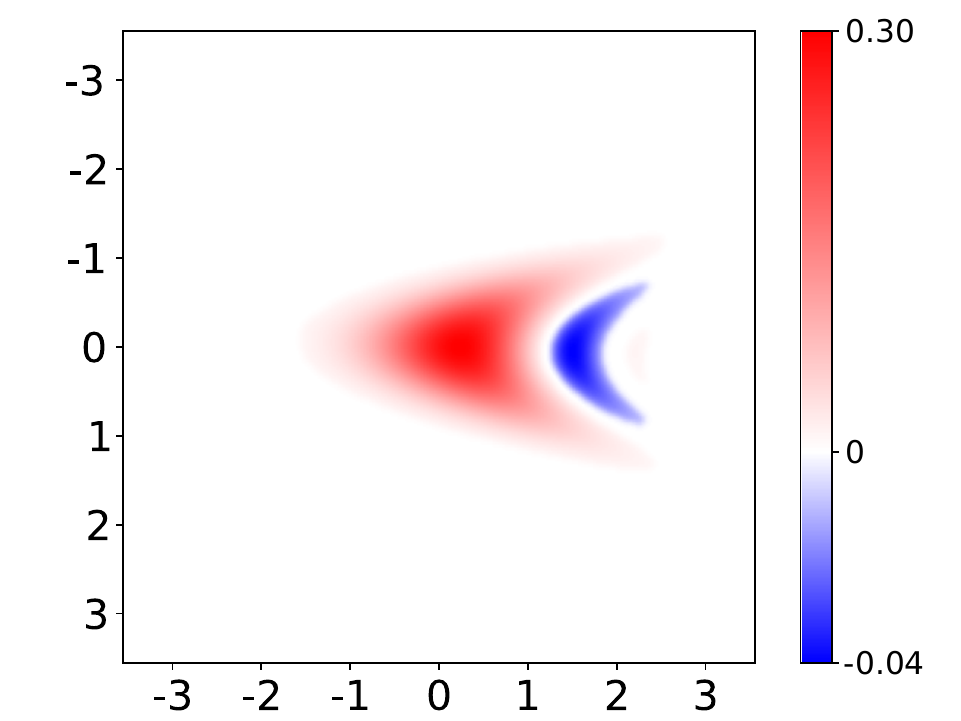}\label{apcube-nxyy}}
    \hskip 5pt
    \subfloat[keep all terms in $\hat H_3'(t)$, add damping $\kappa_a=10\,\mu\mathrm{s}^{-1}$ to SNAIL mode, and displacement drive that cancels \eqref{eq:cubecounter}]{\includegraphics[width=0.26\textwidth]{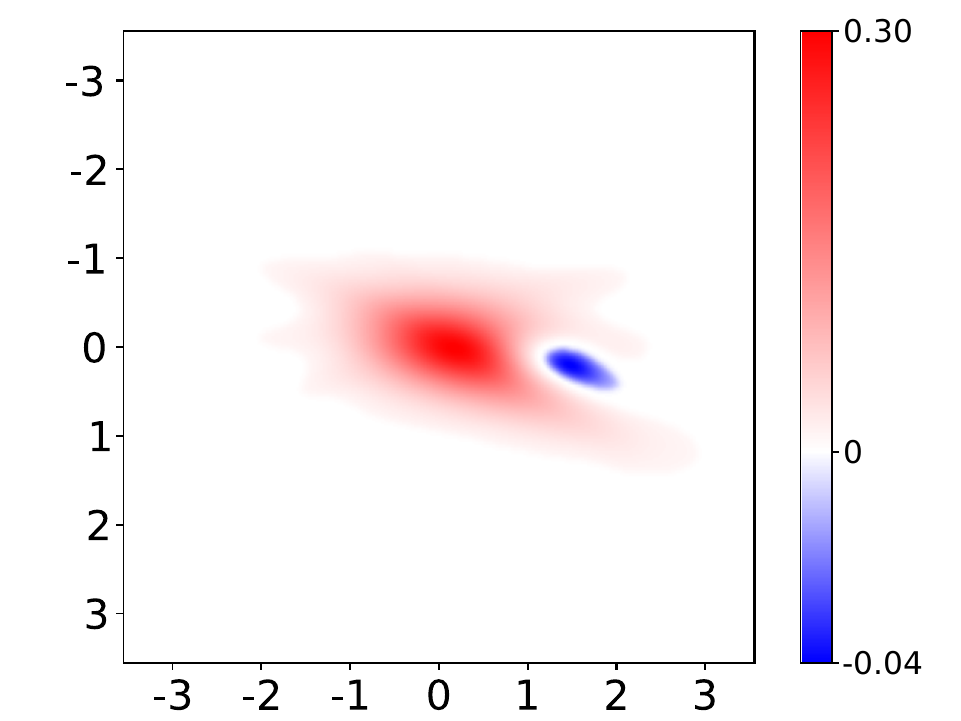}\label{apcube-damp}}
    \caption{\label{apfig-cube}Evolution of $|0\rangle$ under different Hamiltonians, cubic phase drive $\epsilon(t)=\Omega\cos(\omega_ct)+\Omega\cos(3\omega_ct)$, $T=0.5\,\mu\mathrm{s}$.}
\end{figure}

We run the cubic phase gate simulations under the same conditions as above, with $\Omega_1=\Omega_3\equiv\Omega$, except adding a drive to cancel the induced displacement. This is directly the consequence of $\hat X^2\hat Y$ becoming resonant:
\begin{eqnarray}
    -3\lambda\epsilon(t)\hat X^2\hat Y&=&-3\lambda\epsilon(t)(\hat a^\dagger+\hat a)(\hat a^\dagger+\hat a)(\hat c^\dagger+\hat c)\rightarrow-3\lambda\epsilon(t)(\hat c^\dagger+\hat c)\label{eq:a00}\\
    \hat H_{\mathrm{disp}}&=&-\frac{3}{2}\lambda\Omega(\hat c^\dagger e^{-i\omega_c t}+\hat ce^{i\omega_c t})\label{eq:cubecounter}
\end{eqnarray}
The right arrow in Eq.~\eqref{eq:a00} assumes that mode $a$ stays (because of the significant damping) in vacuum so that $\langle(\hat a^\dagger+\hat a)^2\rangle\approx1$. 
Fig.~\ref{apfig-cube} shows the final states we obtain. 
Fig.~\ref{apcube-nxyy} serves as a reference for an ideal cubic phase state, it is characterized by a parabolic squeezing shape~\cite{radimflip}, which is roughly resembled by Fig.~\ref{apcube-damp}. 
Fitting the rotation of these two states gives a maximum fidelity of 0.81. Although this fidelity (0.81) is close to that of trisqueezing (0.83), the shape is reproduced much less faithfully. 
Comparing Figs.~\ref{apfig-wig} and~\ref{apfig-cube}, we see that the implementation of a cubic phase gate is much more demanding (or vulnerable) than trisqueeze.

\clearpage
\newpage

\section{System Dynamics: Floquet}\label{appendix:flo}
In this section, we investigate the dynamics of a driven nonlinear oscillator, namely a SNAIL. This is crucial to the understanding of a SNAIL-HBAR coupled system because the beam-splitter interaction between these two elements is linear, meaning that if one captures the behavior of a driven SNAIL, the HBAR mode is expected to evolve in a similar manner.

The single-mode Hamiltonian of a SNAIL with frequency $\omega\equiv\omega_a^0$, flux-modulated at $\omega_d\simeq3\omega$, is (for brevity, in this and the next section we write $\hat a\equiv\hat a_0$):
\begin{equation}
    \hat H(t)=\omega \hat a^\dagger \hat a+g^{(3)}_{\rm dc}(\hat a^\dagger+\hat a)^3+g^{(4)}_{\rm dc}(\hat a^\dagger+\hat a)^4+\frac{\Omega_3}{2}(e^{i\omega_dt}+e^{-i\omega_dt})(\hat a^\dagger+\hat a)^3\label{eq:g3}
\end{equation}
with $\Omega_3=\tilde g^{(3)}_{\rm ac}\phi_{\rm ac}$ as in the main text. The modulated linear term $g^{(1)}_{\rm ac}$ has been removed by the displacement of Appendix~\ref{appendix:disp}, while the modulated terms $g^{(2)}_{\rm ac}$ and $g^{(4)}_{\rm ac}$ are non-resonant and are dropped. We keep the static quartic term $g^{(4)}_{\rm dc}$, since at the KFP it cancels the Kerr nonlinearity induced by $g^{(3)}_{\rm dc}$. This is a periodic Hamiltonian, which suggests a natural Floquet treatment. 
We typically aim to ramp up adiabatically the drive amplitude $\Omega_3(t)$ from zero, such that the system stays on Floquet state.

To solve the Floquet spectrum $\epsilon_N$, we are going to deploy the Shirley method~\cite{shirley}. 
We separate Eq.~\eqref{eq:g3} into three parts by their frequencies: $\hat H_0=\omega \hat a^\dagger \hat a+g^{(3)}_{\rm dc}(\hat a^\dagger+\hat a)^3+g^{(4)}_{\rm dc}(\hat a^\dagger+\hat a)^4$, $\hat H_1(t)=(\Omega_3/2)e^{i\omega_dt}(\hat a^\dagger+\hat a)^3$, $\hat H_{-1}(t)=(\Omega_3/2)e^{-i\omega_dt}(\hat a^\dagger+\hat a)^3$. Solving $\hat H_0$ to leading order in $g^{(3)}_{\rm dc}$ and $g^{(4)}_{\rm dc}$ gives unperturbed energy $\epsilon_N^{(0)}$, then $\hat H_{\pm 1}(t)$ offer second order correction $\epsilon_N^{(2)}$.
First, we are going to find out unperturbed energy $\epsilon_N^{(0)}$ from $\hat H_0$: $g^{(4)}_{\rm dc}(\hat a^\dagger+\hat a)^4$ gives a first order correction, while $g^{(3)}_{\rm dc}(\hat a^\dagger+\hat a)^3$ is off-diagonal and gives a second order correction to free energy $N\omega$, so that:
\begin{align}
    \epsilon_N^{(0)}&=N\omega+g^{(4)}_{\rm dc}\langle N|(\hat a^\dagger+\hat a)^4|N\rangle+\sum_{M\neq N}\frac{\left|\langle N|g^{(3)}_{\rm dc}(\hat a^\dagger+\hat a)^3|M\rangle\right|^2}{N\omega-M\omega}\nonumber\\
    &=N\omega+g^{(4)}_{\rm dc}\left(6N^2+6N+3\right)+\left(g^{(3)}_{\rm dc}\right)^2\left[\frac{(N+1)(N+2)(N+3)}{-3\omega}+\frac{N(N-1)(N-2)}{3\omega}+\frac{9(N+1)^3}{-\omega}+\frac{9N^3}{\omega}\right]\nonumber\\
    &=N\omega+g^{(4)}_{\rm dc}\left(6N^2+6N+3\right)-\frac{\left(g^{(3)}_{\rm dc}\right)^2}{\omega}(30N^2+30N+11)\label{eq:floquete0}
\end{align}
At the KFP, $12g^{(4)}_{\rm dc}=60(g^{(3)}_{\rm dc})^2/\omega$ (see Appendix~\ref{appendix:kerr}), and Eq.~\eqref{eq:floquete0} reduces to $\epsilon_N^{(0)}=N\omega+4(g^{(3)}_{\rm dc})^2/\omega$, i.e., the levels are equally spaced at this order.
We then construct the Shirley extension formalism, to do that we introduce an auxiliary degree of freedom, which in fact corresponds to the drive photon. For one frequency component $\hat H_m(t)=\hat h_me^{i(m\omega_d)t}$, its extension is given by
\begin{equation}
    \hat H_m(t)\rightarrow\sum_{k}|k-m\rangle(\hat h_m+\delta_{m0}k\omega_d)\langle k| \;, \label{eq:shirleyxpand}
\end{equation}
with $\delta_{m0}$ being the Kronecker delta.
In this way, one eliminates time dependency from system equation and turns it into an eigenvalue problem. We write the full Hamiltonian \eqref{eq:g3}'s Shirley extension as
\begin{align}
    \hat H_{\rm shir}=&\sum_{K}|K\rangle(\hat h_0+K\omega_d)\langle K|\nonumber\\
    +&\sum_{NK}\frac{\Omega_3}{2}\sqrt{(N+3)(N+2)(N+1)}\left(|N+3,K+1\rangle\langle N,K|+|N,K+1\rangle\langle N+3,K|+h.c.\right)\nonumber\\
    +&\sum_{NK}\frac{3\Omega_3}{2}\sqrt{(N+1)^3}\left(|N+1,K+1\rangle\langle N,K|+|N,K+1\rangle\langle N+1,K|+h.c.\right) \;. \label{eq:shirleybraket}
\end{align}
This new Hamiltonian has a block diagonal part $\sum_{K}|K\rangle(\hat h_0+K\omega_d)\langle K|$, whose energy is $\epsilon_{NK}^{(0)}=\epsilon_N^{(0)}+K\omega_d$.
We treat the remaining parts (2nd and 3rd line of Eq.~\eqref{eq:shirleybraket}) in second order perturbation theory. 
Denoting $\Delta_{N+l}\equiv\epsilon_N^{(0)}-\epsilon_{N+l}^{(0)}$, we find $\epsilon_N^{(2)}$ to be
\begin{align}
    \epsilon_N^{(2)}/(\Omega_3/2)^2&=\frac{2\Delta_{N+3}}{\Delta_{N+3}^2-\omega_d^2}(N+3)(N+2)(N+1)+\frac{2\Delta_{N-3}}{\Delta_{N-3}^2-\omega_d^2}N(N-1)(N-2)\nonumber\\
    &+\frac{18\Delta_{N+1}}{\Delta_{N+1}^2-\omega_d^2}(N+1)^3+\frac{18\Delta_{N-1}}{\Delta_{N-1}^2-\omega_d^2}N^3 \;. \label{eq:floquete2}
\end{align}
Combining Eq.~\eqref{eq:floquete0} and \eqref{eq:floquete2} we get an approximate expression for the Floquet quasienergies $\epsilon_N$. For weak drive, $\Omega_3\ll|\omega_d-3\omega|$, the drive-induced shifts of all levels follow a parabolic rule $\epsilon_N-\epsilon_N^{(0)}\propto \Omega_3^2$.

\begin{figure}[h]
    \centering
    \includegraphics[width=0.8\textwidth]{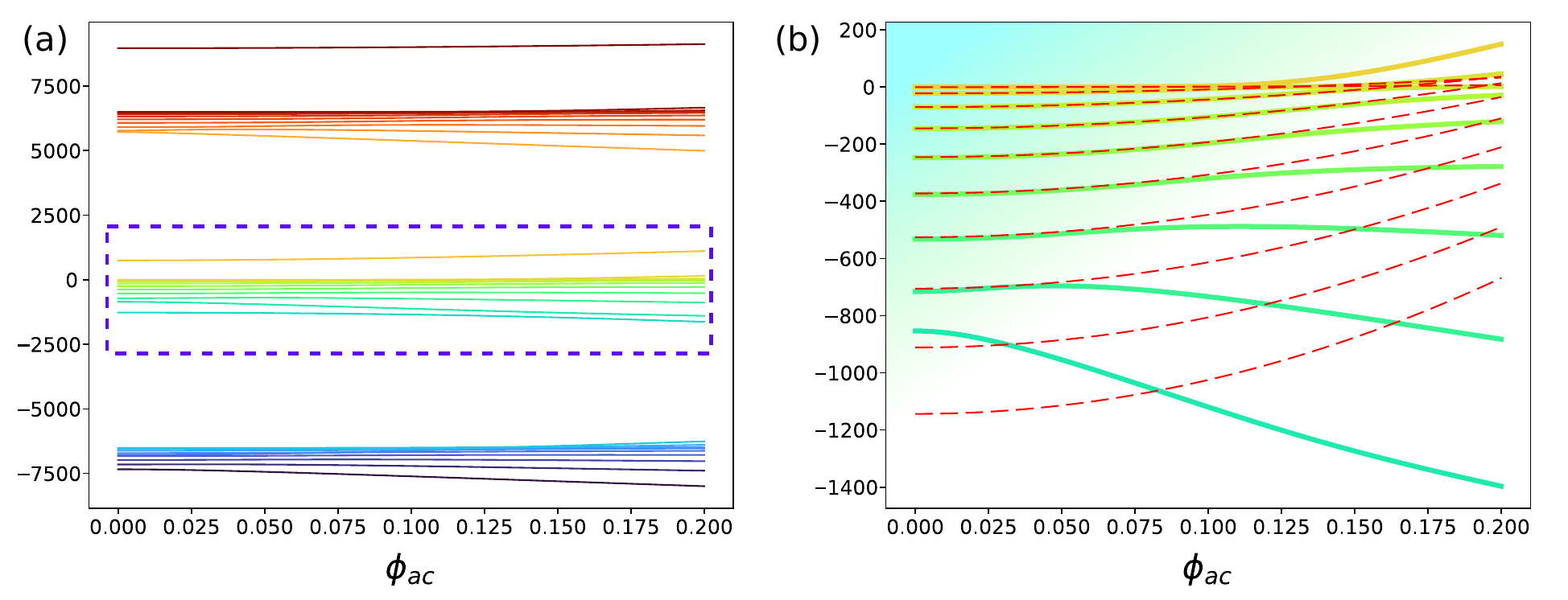}
    \caption{(a) Floquet spectrum of a driven SNAIL oscillator whose Hilbert space truncation is set to 30. From the data obtained through diagonalization of $\hat U(T)$, three distinct groups near $0,\pm\omega_d/3$ are clearly visible. (b) Zoom-in view of the middle group in (a) (indicated by purple dashed box). Red dashed lines are the $\epsilon_N$'s calculated through Shirley extension, showing that the two methods agree well inside teal shaded region (top left corner), where only lower Floquet levels are present (not affected by Hilbert space truncation) and $\phi_{\rm ac}$ is small (perturbation theory holds).}
    \label{apfig-floq}
\end{figure}

Fig.~\ref{apfig-floq} shows the calculated Floquet quasienergy levels of Eq.~\eqref{eq:g3}. 
When the drive amplitude $\phi_{\rm ac}$ is small, the lowest levels (in absolute value) obtained through direct diagonalization of time evolution operator $\hat U(T)$ agree well with our calculated $\epsilon_N=\epsilon_N^{(0)}+\epsilon_N^{(2)}$, which is a family of parabolas. 
Higher levels show relatively larger discrepancies because of Hilbert space truncation, while in the regions where $\phi_{\rm ac}$ becomes bigger, perturbation theory no longer applies.

\begin{figure}[h]
    \centering
    \includegraphics[width=0.92\textwidth]{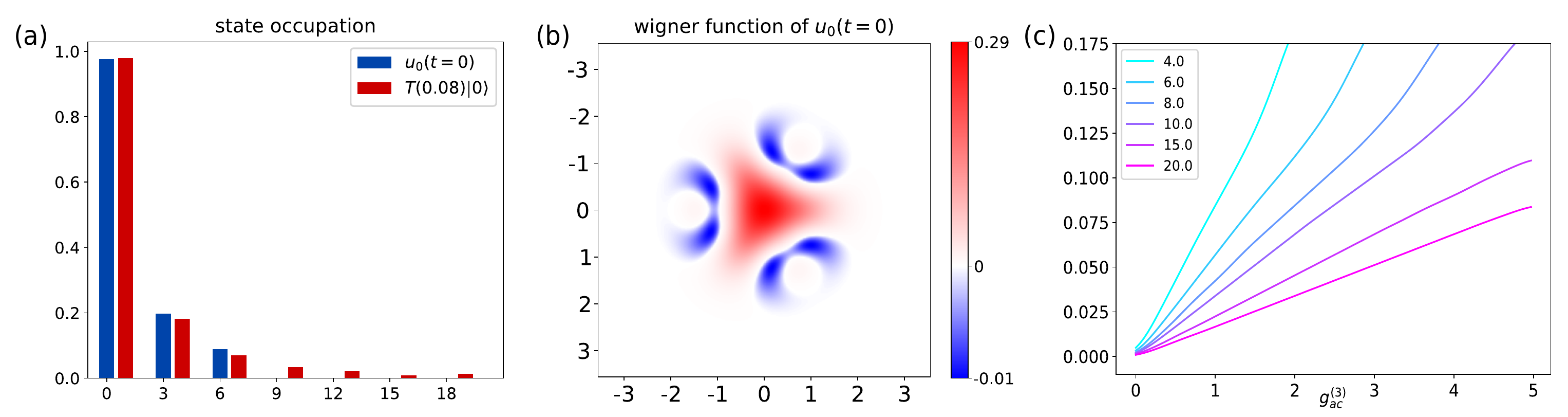}
    \caption{Floquet ground state $|u_0(0)\rangle$ under $(\omega,\omega_d,\Omega_3)=(6500,19530,5.0)$\unit{}{MHz}, i.e., $\omega_d-3\omega=\unit{30}{MHz}$, its closest ideal state has $C=0.08$. (a) Fock state occupation, red bars (ideal state) are shifted one unit to the right for clarity. (b) Wigner function of $|u_0(0)\rangle$. (c) For different drive detunings $\omega_d-3\omega$, closest ideal state parameter $C_{\rm ideal}$ is plotted against the drive amplitude $\Omega_3/2$.}
    \label{apfig-floccupation}
\end{figure}

Next, we will search for the Floquet ground state directly from the Shirley Hamiltonian \eqref{eq:shirleybraket}. 
Solving it to first order gives $|3\rangle$ state occupation, while to second order gives $|6\rangle$ state occupation, which are the two main components of a trisqueezed state besides vacuum $|0\rangle$. 
Eigenvectors $|\phi\rangle$ of Eq.~\eqref{eq:shirleybraket} are in the product space $|N,K\rangle$, separating its $K$ component will give us a set of vectors $\{|u(k)\rangle\}$ in the original $N$ space through
% \mf{here and below, you cannot have KETS on the right hand side and no kets on the LHS...} \yz{checked}
\begin{equation}
|\phi\rangle=\sum_{n,k}c_{n,k}|n,k\rangle=\sum_{k}\left(\sum_{n}c_{n,k}|n\rangle\right)|k\rangle\equiv\sum_{k}|u(k)\rangle|k\rangle , \label{eq:uk}
\end{equation}
where $|u(k)\rangle$ are the Fourier coefficients of the Floquet state $|u(t)\rangle$
\begin{equation}
    |u(t)\rangle=\sum_{k}|u(k)\rangle e^{ik\omega_d t} \;. \label{eq:utk}
\end{equation}
Let us denote the off-diagonal parts of \eqref{eq:shirleybraket} as $\hat V$, the first order correction to the unperturbed eigenstate $|N=0,K=0\rangle$ is:
\begin{equation}
    |\phi_{0,0}^{(1)}\rangle=\sum_{n,k}\frac{\langle n,k|\hat V|0,0\rangle}{-n\omega-k\omega_d}|n,k\rangle.
\end{equation}
There are in fact four combinations of $(n,k)$ that have non-zero matrix element, namely $(n,k)=(1,\pm1),(3,\pm1)$, but we are mainly interested in the dominant one.
It is straightforward to see that, since the drive frequency $\omega_d$ is set to be near resonant $\omega_d\simeq 3\omega$, the dominant term we are seeking for is $(n,k)=(3,-1)$. 
This gives the Fourier component $|u_0(k=-1)\rangle$ that contains mainly $|n=3\rangle$:
\begin{equation}
    |u_0(k=-1)\rangle=\frac{\sqrt{6}\,\Omega_3}{2(\omega_d-3\omega)}|3\rangle.\label{eq:floquetp1}
\end{equation}
The same can be done for the second order correction, which has dominant term $(n_1,k_1,n_2,k_2)=(6,-2,3,-1)$.
\begin{eqnarray}
    |\phi_{0,0}^{(2)}\rangle=\sum_{n_1,k_1}|n_1,k_1\rangle\sum_{n_2,k_2}\frac{\langle n_1,k_1|\hat V|n_2,k_2\rangle\langle n_2,k_2|\hat V|0,0\rangle}{(-n_1\omega-k_1\omega_d)(-n_2\omega-k_2\omega_d)},\\
    \Rightarrow |u_0(k=-2)\rangle=\frac{3\sqrt{5}\,\Omega_3^2}{2(\omega_d-3\omega)^2}|6\rangle.\label{eq:floquetp2}
\end{eqnarray}
Combining Eqs.~(\ref{eq:utk}, \ref{eq:floquetp1}, \ref{eq:floquetp2}), we get the Floquet solution for a trisqueezed state. This solution is valid when $|\omega_d-3\omega|\gg \Omega_3$ and it gives
\begin{equation}
    |u_0(t)\rangle=|0\rangle+\frac{\sqrt{6}\,\Omega_3}{2(\omega_d-3\omega)}|3\rangle e^{-i\omega_dt}+\frac{3\sqrt{5}\,\Omega_3^2}{2(\omega_d-3\omega)^2}|6\rangle e^{-2i\omega_dt}.\label{eq:trisqfloquet}
\end{equation}
To quantify how good this Floquet state resembles an ideal trisqueeze state $|\psi(C)\rangle=\hat T(C)|0\rangle$, with $\hat T(C)$ as defined in Eq.~\eqref{trisqueeze} of the main text, we look for a parameter $C_{\rm ideal}$ that has highest fidelity $\mathcal{F}(|u_0(t)\rangle,\hat T(C_{\rm ideal})|0\rangle)$, normally we take $t=0$. In Fig.~\ref{apfig-floccupation}(a)(b) we show an example of Floquet state and compare it to the closest ideal state. 
Although higher number states starting from $|9\rangle$ are omitted in $|u_0\rangle$ due to Hilbert space truncation, the result still yields a 99.87\% fidelity.
\\Fig.~\ref{apfig-floccupation}(c) shows the relationship between the drive amplitude $\Omega_3$ and $C_{\rm ideal}$, each colored line indicates a different detuning value. We can see $C_{\rm ideal}$ is roughly linear under relatively small $\Omega_3$ and that it scales up quickly when bringing the drive frequency close to resonance. 
It is worth pointing out that one should avoid driving too close to resonance, as Eq.~\eqref{eq:trisqfloquet} will break down. The exact solution of our Floquet state also undergoes a population inversion between $|0\rangle$ and $|3\rangle$ state, meaning it is no longer a trisqueeze state.

\clearpage
\newpage

\section{System Dynamics: Anti-Cross}\label{appendix:trans}
In this section we investigate the effect of strong flux driving, but still in the adiabatic ramp-up case. 
As already seen in Fig.~\ref{apfig-floq}(b), the system dynamics is governed by Floquet modes, which under low drive powers are well separated.
Increasing the drive power will cause Floquet modes to anti-cross with each other, such that tunneling can occur.
\begin{figure}[h]
    \centering
    \includegraphics[width=0.5\textwidth]{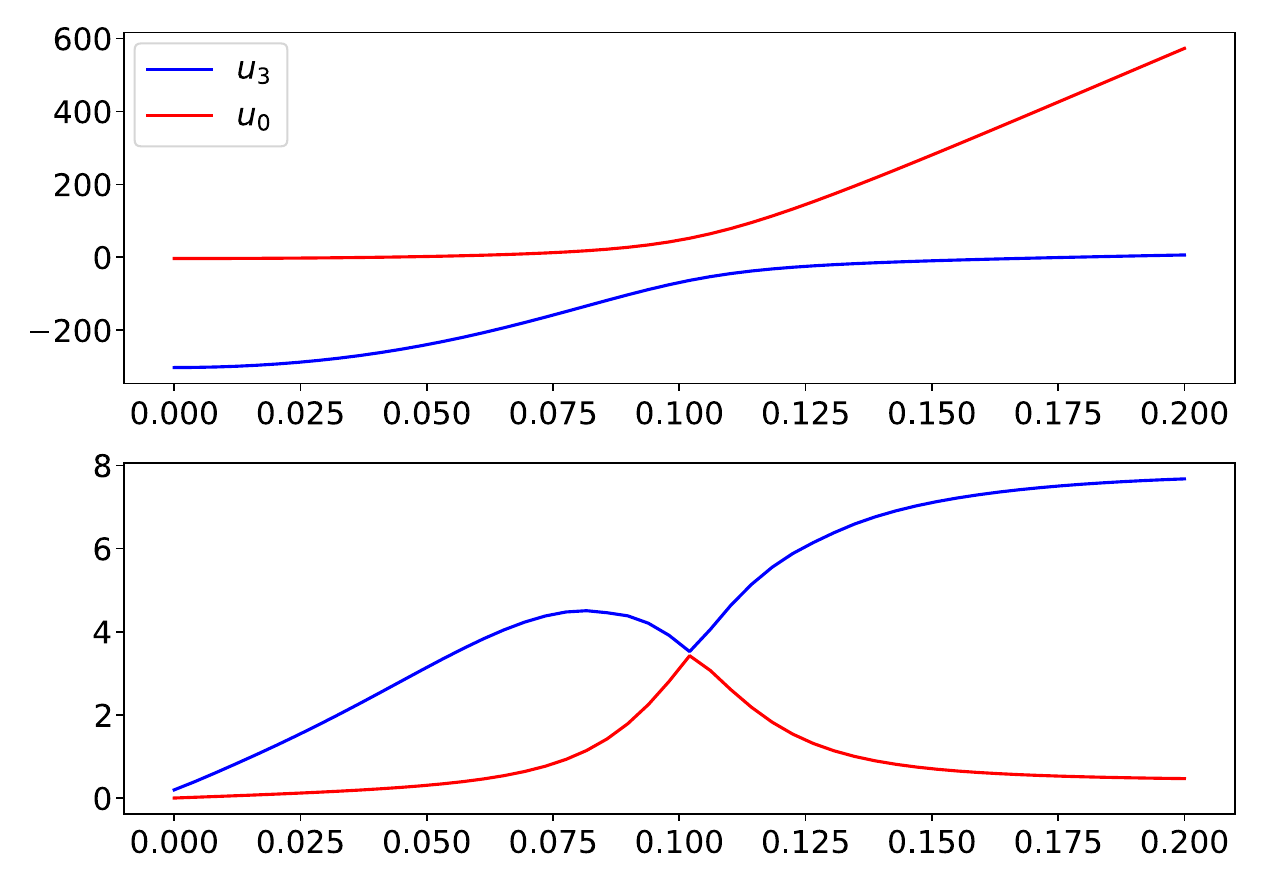}
    \caption{Top: two lowest modes $u_0$ and $u_3$ of the Floquet spectrum, anti-crossing around flux amplitude 0.10 is observed. Bottom: skewness of the modes $u_{0,3}$, from where the anti-crossing can also be identified.}
    \label{apfig-anticrosssk}
\end{figure}

As an example, Fig.~\ref{apfig-anticrosssk} shows the two lowest Floquet modes. The target trisqueeze state we expect to obtain from Floquet dynamics \eqref{eq:trisqfloquet} is the Floquet mode connected to vacuum state $|0\rangle$ (red curve), but after its first anti-crossing with another mode $u_m(t)$ which originates from a higher number state $|m\rangle$ (in our case $m=3$, blue curve), their profile exchanges, so that we will continue with $u_m(t)$ along adiabatic path. 
One can get an even clearer image in skewness profile of the two modes, this explains the phenomenon we observed in main text Fig.~\ref{skewcomp}(b), where increasing flux drive power resulted in a counterintuitive decrease of skewness.

We can derive from \eqref{eq:floquete2} an approximate expression for the anti-crossing. Substituting $N=0,3$, we get
\begin{equation}
    \epsilon_0\simeq\frac{3\,\Omega_3^2}{2(\omega_d-3\omega)}\qquad \epsilon_3\simeq-(\omega_d-3\omega)+\frac{57\,\Omega_3^2}{2(\omega_d-3\omega)}
\end{equation}
Setting $\epsilon_0=\epsilon_3$, one can solve for the point $\Omega_{3,\rm cross}$ where the two levels anti cross, then use the relation $\Omega_3=\tilde g^{(3)}_{\rm ac}\phi_{\rm ac}$, with $\tilde g^{(3)}_{\rm ac}=E_J\tilde c_3'\varphi_{\rm zpf}^3/6$ [see Eq.~\eqref{definition:gn} and Appendix~\ref{appendix:disp}], to get the corresponding flux drive amplitude $\phi_{\rm ac,cross}$ as 
\begin{equation}
    \Omega_{3,\rm cross}=\frac{\omega_d-3\omega}{\sqrt{27}}\qquad \phi_{\rm ac,cross}=\frac{6\,\Omega_{3,\rm cross}}{E_J\tilde c_3'\varphi_{\rm zpf}^3}=\frac{2}{\sqrt{3}\,E_J\tilde c_3'\varphi_{\rm zpf}^3}(\omega_d-3\omega) \,. \label{appendix:anticross-spot}
\end{equation}
Using the values for the simulation in Fig.~\ref{apfig-anticrosssk}, Eq.~\eqref{appendix:anticross-spot} gives $\phi_{\rm ac,cross}=0.1027$, consistent with Fig.~\ref{apfig-anticrosssk}. 

\begin{figure}[t]
    \centering
    \includegraphics[width=0.7\textwidth]{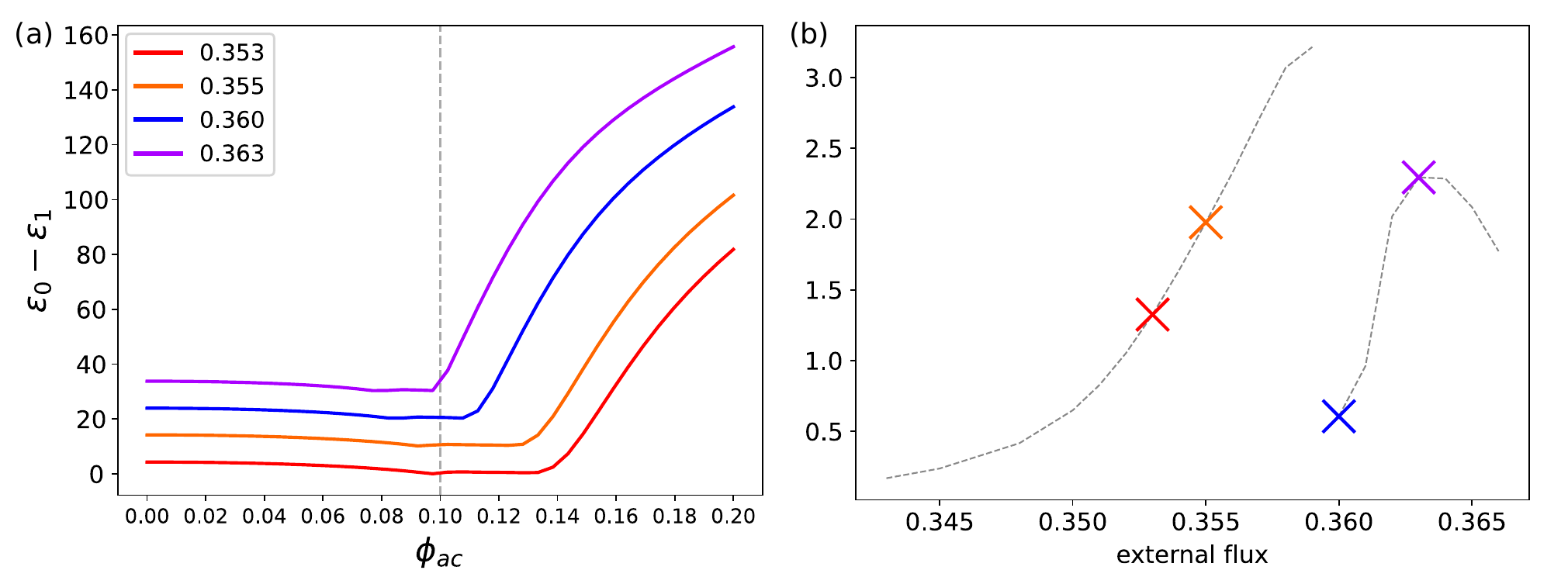}
    \caption{(a) Quasienergy gap between the lowest two Floquet modes, where a turning point indicates anti-crossing. The flux amplitude is fixed at 0.1 (grey vertical line) and curves are artificially displaced vertically for visual clarity. (b) Same plot as Fig.~\ref{skewcomp}(a), where bias points referring to 4 distinct bands in (a) are highlighted in their respective colors.}
    \label{apfig-passive}
\end{figure}

At the same time, Fig.~\ref{skewcomp}(a) of the main text can be explained by the notion of relative movement. When we change the flux bias $\Phi_{\rm dc}$, the Floquet spectrum moves, so that we go through anti-crossings at larger $\Phi_{\rm dc}$'s while staying away from anti-crossings at smaller $\Phi_{\rm dc}$'s. 
Fig.~\ref{apfig-passive} shows anti-crossing of the lowest two Floquet modes (one being the trisqueeze mode $u_0$) moves past $\phi_{\rm ac}=0.10$, from right to left, as a function of increasing $\Phi_{\rm dc}/\Phi_0$ from 0.353 to 0.363. 

\begin{figure}[t]
    \centering
    \includegraphics[width=0.6\textwidth]{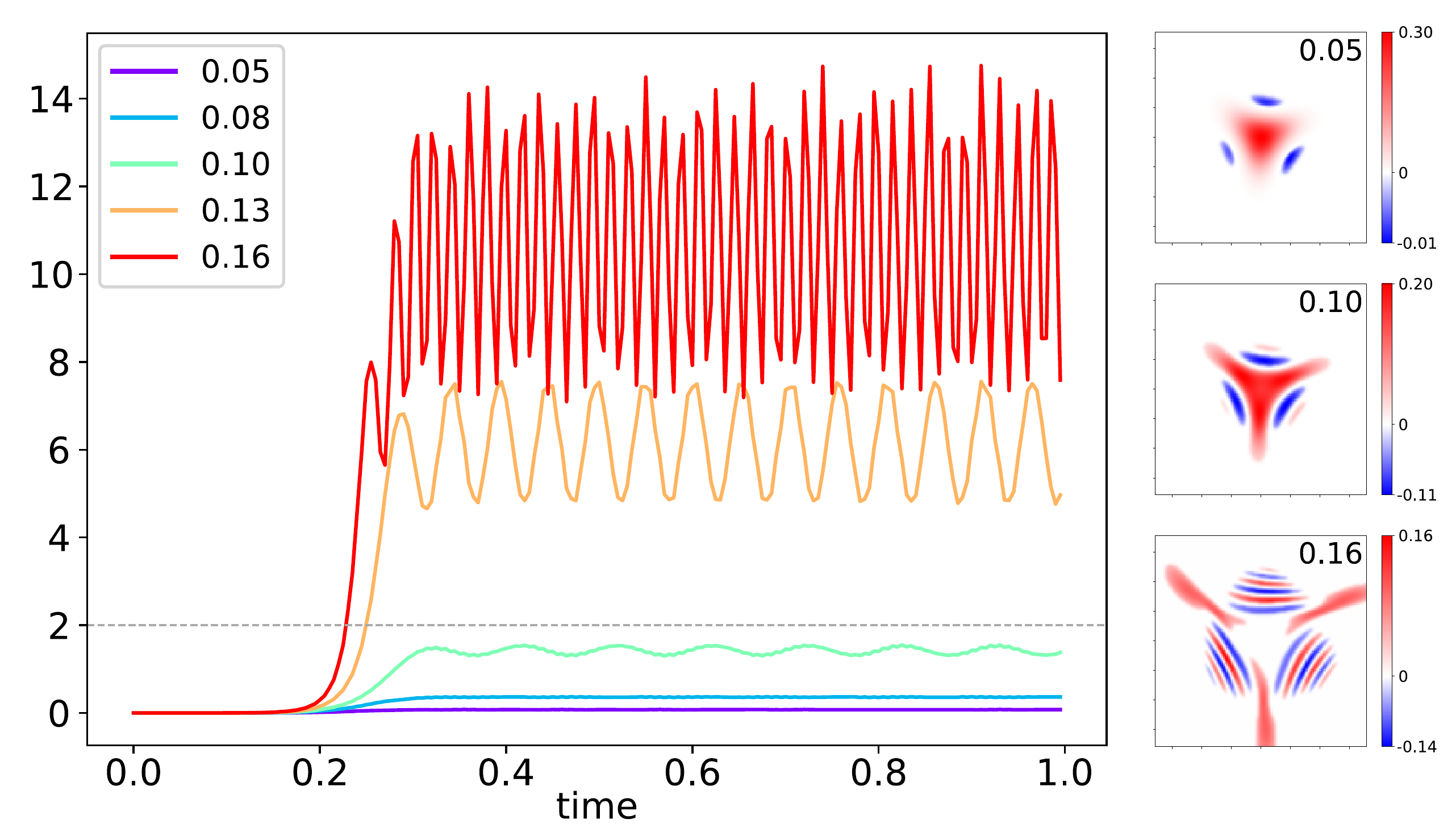}
    \caption{Average excitation number $\langle \hat a^\dagger \hat a\rangle$ of SNAIL mode over time during evolution. Colored curves correspond to different flux drive amplitudes ranging from 0.05 to 0.16. Grey dotted line indicates $\langle \hat a^\dagger \hat a\rangle$ for ideal trisqueeze state $\hat T(0.2)|0\rangle$, while subplots on the right show Wigner function of the final state for three $\phi_{\rm ac}$ values $(0.05,0.10,0.16)$.}
    \label{apfig-anticross}
\end{figure}

To see how states are affected in a time dependent simulation, we adiabatically ramp up the parametric drive during the first \unit{0.2}{\mu s}, after which its amplitude stays constant at its maximum value. We track average excitation count on the SNAIL mode, as well as the final state at $t=\unit{1}{\mu s}$. 
We can see from Fig.~\ref{apfig-anticross} that flux amplitudes smaller than 0.10 produce well shaped trisqueeze state, the grey dotted line, corresponding to $\hat T(0.2)|0\rangle$ computed in the same truncated Fock space, serves as a rough estimate of the achievable trisqueezing upper bound and it is above all three curves at the bottom. Overshooting the grey line suggests a major deviation from the ideal trisqueezing dynamics, as higher number states are populated. This is the case for $\phi_{\rm ac}$ values $(0.13,0.16)$. Again, the turning point is slightly bigger than 0.10, close to the predicted value 0.1027.

%This is also a thing to note in numerical simulations, because of Hilbert space truncation, one should always avoid exciting the uppermost few states, or otherwise numerical results won't be credible.

\clearpage
\newpage

\section{Circuit Quantization of SNAIL-HBAR-Transmon Coupled System}\label{appendix:cqed}
In this section we derive the quantized circuit Hamiltonian in main text Eq.~\eqref{cqed1}. For simplicity, we set here $\hbar=1$.
\begin{figure}[h]
    \includegraphics[width=0.4\textwidth]{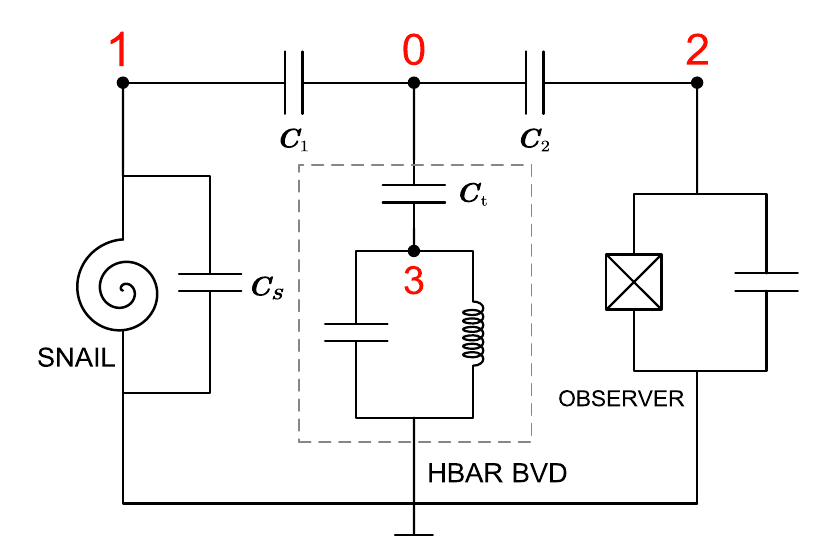}
    \caption{\label{circuitappend}Circuit diagram of the proposed SNAIL-HBAR-Transmon system.}
\end{figure}
\\In Fig.~\ref{circuitappend} we label four circuit nodes $i=0,1,2,3$ (dummy, SNAIL, transmon, HBAR in order), and take their associated node fluxes as $\Phi_i$. 
As we are applying the ``black box quantization'' method, we temporarily ignore nonlinear inductances in SNAIL and spectator qubit. We write the system's capacitive energy $H_c$ in matrix form
\begin{equation}
    H_c=\frac{1}{2}\mathbf{\dot{\Phi}^T}
    \begin{bmatrix}
        C_1+C_2+C_t & -C_1 & -C_2 & -C_t \\
        -C_1 & C_1+C_s & 0 & 0 \\
        -C_2 & 0 & C_2+C_q & 0 \\
        -C_t & 0 & 0 & C_m+C_t \\
    \end{bmatrix}\mathbf{\dot{\Phi}}\equiv\frac{1}{2}\mathbf{\dot{\Phi}^T}\mathbf{C}\mathbf{\dot{\Phi}}
    \qquad
    \mathbf{\dot{\Phi}}\equiv
    \begin{bmatrix}
        \dot{\Phi}_0\\
        \dot{\Phi}_1\\
        \dot{\Phi}_2\\
        \dot{\Phi}_3\\
    \end{bmatrix}
    \label{cqedmatrix} \;.
\end{equation}
Note that node 0 (dummy) is not a free node, since it has to satisfy charge preservation $Q_0=0$. Because $H_c$ is quadratic in $\dot{\Phi}_i$, finding conjugate momenta $Q_i=\frac{\partial L}{\partial\dot{\Phi}_i}$ and re-expressing \eqref{cqedmatrix} in terms of the $Q_i$ results in
\begin{equation}
    H_c=\frac{1}{2}\mathbf{Q^T}\mathbf{C^{-1}}\mathbf{Q}=\frac{1}{2}\mathbf{Q^T}\frac{\mathbf{B}}{\det(\mathbf{C})}\mathbf{Q}
    \qquad
    \mathbf{Q}\equiv \left[Q_0,Q_1,Q_2,Q_3\right]^T ,
    \label{cqedc-1}
\end{equation}
where $\mathbf{B}$ is the cofactor matrix and $\det(\mathbf{C})$ stands for the determinant. Here, applying charge constraint $Q_0=0$ is equivalent to removing 1st row and 1st column in matrix representation, so that \eqref{cqedc-1} can be explicitly written as:
\begin{equation}
    H_c=\frac{1}{2\det(\mathbf{C})}\mathbf{Q^T}
    \begin{bmatrix}
        B_{11} & C_1C_2(C_m+C_t) & C_1C_t(C_2+C_q) \\
        C_1C_2(C_m+C_t) & B_{22} & C_2C_t(C_1+C_s) \\
        C_1C_t(C_2+C_q) & C_2C_t(C_1+C_s) & B_{33}
    \end{bmatrix}
    \mathbf{Q}
    \qquad
    \mathbf{Q}\equiv
    \begin{bmatrix}
        Q_1\\
        Q_2\\
        Q_3\\
    \end{bmatrix}
    \label{cqedq}
\end{equation}
\begin{align}
B_{11}&=C_1(C_2+C_q)(C_m+C_t)+C_2C_q(C_m+C_t)+(C_2+C_q)C_mC_t\\
B_{22}&=C_2(C_1+C_s)(C_m+C_t)+C_1C_s(C_m+C_t)+(C_1+C_s)C_mC_t\\
B_{33}&=C_t(C_1+C_s)(C_2+C_q)+C_1C_s(C_2+C_q)+(C_1+C_s)C_2C_q
\end{align}
Off-diagonal elements $B_{ij}$ give couplings between different nodes. Substituting $\Phi_i\rightarrow\phi_i/2e$ and $Q_i\rightarrow2e\cdot n_i$ gives the dimensionless variables $\phi_i$ and $n_i$. At the same time, we define the mode charging energy
\begin{equation}
    E_{c,i}=\frac{e^2B_{ii}}{2\det(\mathbf{C})} \;. \label{apec}
\end{equation}
We can now write all the diagonal terms, including both linear and nonlinear parts of the inductive energy, as
\begin{equation}
    H_{\mathrm{diag}}=H_I+H_c+H'=\sum_i \frac{1}{2}E_{J,i}\,(\phi_i-\phi_\mathrm{offset,i})^2+4E_{c,i}\,n_i^2+H'(\phi_i) . \label{apejec}
\end{equation}
Here, $H'(\phi_i)$ are the previously omitted nonlinear energies, and $E_{J,i}$ are the linear inductive energies: $E_{J,1}=E_Jc_2$ for the SNAIL at its bias point, $E_{J,2}$ for the linearized transmon junction, and $E_{J,3}=(\hbar/2e)^2/L_m$ for the mechanical inductance (i.e., $hE(L_m)$ in the notation of Table~\ref{cqedtable}). Offset flux $\phi_\mathrm{offset,i}$ is tailored to the SNAIL device, because the SNAIL's potential minimum is not at zero, but $\phi_\mathrm{offset,1}=\phi_m$. One can simply get rid of this offset by substituting $\phi_1-\phi_m\rightarrow\phi_1$, while the charge $n_1$ is unaffected. 
Finally, we perform canonical quantization by turning $\phi_j$ and $n_j$ into operators satisfying the canonical commutation relation $\left[\hat{\phi}_j,\hat{n}_j\right]=i$, $j=1,2,3$. Afterwards, we can reintroduce nonlinearities of both SNAIL and transmon to the model.

Equation~\eqref{apejec} takes the form of a set of free harmonic oscillators. Expressing $\hat{\phi}_j$ and $\hat{n}_j$ in ladder operators and zero point fluctuations
\begin{equation}
    \hat{\phi}_j=\phi_{0,j}(\hat a_j^\dagger+\hat a_j)\;, \qquad\hat{n}_j=i n_{0,j}(\hat a_j^\dagger-\hat a_j) . \label{apladder}
\end{equation}
Upon substitution, the quantized $\hat H_{\mathrm{diag}}$ becomes the familiar form $\sum_i \omega_i \hat a_i^\dagger \hat a_i$ with bare frequencies given by $\omega_i=\sqrt{8E_{J,i}E_{c,i}}$. 
Coupling terms in Eq.~\eqref{cqedq} are now
\begin{align}
    \hat V_{ij}=-\frac{4e^2 B_{ij}}{\det(\mathbf{C})}n_{0,i}n_{0,j}(\hat a_i^\dagger-\hat a_i)(\hat a_j^\dagger-\hat a_j)\approx g_{ij}(\hat a_i^\dagger \hat a_j+\hat a_i \hat a_j^\dagger)\label{apvij}\\
    g_{ij}=\frac{4e^2 B_{ij}}{\det(\mathbf{C})}n_{0,i}n_{0,j}=\frac{\sqrt{\omega_i\omega_j}}{2}\frac{B_{ij}}{\sqrt{B_{ii}B_{jj}}}\label{apgij}
\end{align}
The approximate sign means we have applied rotating wave approximation so that only beam-splitter interaction is kept. Identifying $(\hat a_1,\hat a_2,\hat a_3)\equiv(\hat a_0,\hat q,\hat c_0)$, $(\omega_1,\omega_2,\omega_3)\equiv(\omega_a,\omega_q,\omega_c)$ and $(g_{12},g_{23},g_{13})\equiv(g_{\rm sq},g_{\rm qm},g_{\rm sm})$, and combining Eq.~\eqref{apvij} with $\hat H_{\mathrm{diag}}$, one recovers the three-mode coupled Hamiltonian in main text, Eq.~\eqref{cqed1}.

\end{document}